\documentclass[aps,prd,twocolumn,showpacs,preprintnumbers,amsmath,amssymb]{revtex4}
\usepackage{graphicx}
\usepackage{dcolumn}
\usepackage{bm}
\usepackage{amssymb,amsmath}
\usepackage{latexsym}

\allowdisplaybreaks

\begin{document}

\title{Displacement-noise-free gravitational-wave detection
with a single Fabry-Perot cavity: a toy model}
\author{Sergey P. Tarabrin and Sergey P. Vyatchanin}
\affiliation{Faculty of Physics, Moscow State University, Moscow,
119992, Russia} \email{tarabrin@phys.msu.ru}
\date{\today}

\begin{abstract}
We propose a detuned Fabry-Perot cavity, pumped through both the
mirrors, as \textit{a toy model} of the gravitational-wave (GW)
detector partially free from displacement noise of the test masses.
It is demonstrated that the noise of cavity mirrors can be
eliminated, but the one of lasers and detectors cannot. The
isolation of the GW signal from displacement noise of the mirrors is
achieved in a proper linear combination of the cavity output
signals. The construction of such a linear combination is possible
due to the difference between the reflected and transmitted output
signals of detuned cavity. We demonstrate that in low-frequency
region the obtained displacement-noise-free response signal is much
stronger than the $f^3_{\textrm{gw}}$-limited sensitivity of
displacement-noise-free interferometers recently proposed by S.
Kawamura and Y. Chen. However, the loss of the resonant gain in the
noise cancelation procedure results is the sensitivity limitation of
our toy model by displacement noise of lasers and detectors.
\end{abstract}
\preprint{LIGO-P070109} 

\maketitle

\section{Introduction}\label{sec_intro}
Currently the search for gravitational radiation from astrophysical
sources is conducted with the first-generation Earth-based laser
interferometers \cite{1972_EMW_detector,2000_GW_detection} (LIGO in
USA \cite{1992_LIGO,2006_LIGO_status,website_LIGO}, VIRGO in Italy
\cite{2006_VIRGO_status,website_VIRGO}, GEO-600 in Germany
\cite{2006_GEO-600_status,website_GEO-600}, TAMA-300 in Japan
\cite{2005_TAMA-300_status,website_TAMA-300} and ACIGA in Australia
\cite{2006_ACIGA_status,website_ACIGA}). The development of the
second-generation GW detectors (Advanced LIGO in USA
\cite{2002_Adv_LIGO_config,website_Adv_LIGO}, LCGT in Japan
\cite{2006_LCGT_status}) is underway.

The sensitivity of the first-generation detectors is limited by a
great amount of noises of various nature: seismic and
gravity-gradient noise at low frequencies (below $\sim 50$ Hz),
thermal noise in suspensions, bulks and coatings of the mirrors
($\sim 50\div 500$ Hz), photon shot noise (above $\sim 500$ Hz),
etc. It is expected that the sensitivity of the second-generation
detectors will be limited by the noise of quantum nature arising due
to Heisenberg's uncertainty principle: the more precise is the
measurement of the test mass coordinate, the more disturbed becomes
its momentum which in turn evolves into the disturbance of the
coordinate, thus ultimately limiting the sensitivity
\cite{1992_quant_meas}. The optimum between measurement noise
(photon shot noise) and back-action noise (radiation pressure noise)
is called the Standard Quantum Limit (SQL)
\cite{1968_SQL,1975_SQL,1977_SQL}.

Though the start of operation of the second-generation detectors is
planned for the next decade, theoretical investigations of the
third-generation prototypes have already begun
\cite{2002_conversion,2002_optical_lever_intracavity,2002_stroboscopic,
2003_optical_lever_bar,2006_intracavity,2007_intracavity}. It is
expected that the barrier of SQL will be overcome and the
sensitivity of the third-stage detectors will be at least an order
of magnitude better than the SQL of a free mass.

Recently in a series of papers
\cite{2004_DNF_GW_detection,2006_DTNF_GW_detection,2006_interferometers_DNF_GW_detection}
S. Kawamura and Y. Chen proposed several topologies of the GW
detectors, both ground- and space-based, which are free from
displacement noise of the test masses --- the noise produced by
external fluctuative forces. For the ground-based optical
interferometers this implies the insusceptibility to seismic noise,
thermal noise in suspensions of the mirrors, etc. However, the most
intriguing feature of displacement-noise-free interferometry (DFI)
is the straightforward overcoming of SQL (since radiation pressure
noise is canceled) without the need of implementation of very
complicated and vulnerable schemes for Quantum-Non-Demolition (QND)
measurements
\cite{1981_squeezed_light,1982_squeezed_light,2002_conversion,1996_QND_toys_tools}.
One only needs to increase laser power to suppress quantum shot
noise and achieve the arbitrarily high sensitivity.

The isolation of the GW signal from fluctuating displacements of the
test masses in the DFI schemes proposed by S. Kawamura {\textit{et
al.}} is possible due to the fact that the interaction of GWs with a
laser interferometer is distributed, as viewed from both the
transverse-traceless (TT) gauge
\cite{1973_gravitation,2003_clas_phys_GW,2005_basics_of_gw_theory}
and the local Lorentz (LL) gauge
\cite{2003_clas_phys_GW,2005_local_observ,2007_GW_FP_LL}.

In the TT gauge test masses are immovable, i.e. have fixed spacial
coordinates and thus do not sense the gravitational wave. However,
GW couples to the light wave in this gauge producing a non-vanishing
phase shift. This can be thought of as an apparent change of the
coordinate velocity of light. Even if the test masses are not
ideally inertial and follow non-geodesic motion then the
interferometer will respond differently to the test masses motions
and the gravitational wave. This difference allows the cancelation
of displacement noise in a proper linear combination of the
interferometer response signals.

From the viewpoint of local observer (the LL gauge) interaction of
the GW with a laser interferometer adds up to two effects. The first
one is the motion of the test masses in the GW tidal force-field. In
this aspect GWs are indistinguishable from any non-GW forces since
both are sensed by the light wave only at the moments of reflection
from the test masses. If the linear scale $L$ of a GW detector is
much smaller than the gravitational wavelength
$\lambda_{\textrm{gw}}$ (the so-called long-wave approximation) then
the effect of the GW force-field is of the order of
$h(L/\lambda_{\textrm{gw}})^0$, where $h$ is the absolute value of
the GW amplitude. Relative motion of the test masses, separated by a
distance $L$, in any force field cannot be sensed by one of them
faster than $L/c$, thus resulting in the rise of terms of the order
of $O[h(L/\lambda_{\textrm{gw}})^1]$ describing time delays which
take the light wave to travel between the masses. Second, GW
directly couples to the light wave effectively changing the
coordinate velocity of light (but in a different manner as compared
to the TT gauge). In long-wave approximation this effect has the
order of $O[h(L/\lambda_{\textrm{gw}})^2]$. Therefore, from the
viewpoint of local observer displacement-noise-free interferometry
necessarily implies the cancelation of non-GW forces along with the
GW force-field leaving a non-vanishing information about the direct
coupling of the GW to light.

It was pointed in Refs.
\cite{2006_DTNF_GW_detection,2006_interferometers_DNF_GW_detection}
that in order the GW detector to be a truly displacement-noise-free
interferometer it should be also free from optical laser noise which
is indistinguishable from laser displacement noise. Cancelation of
the optical noise in interferometric experiments is usually achieved
by implementing the differential (balanced) schemes of measurements:
in conventional interferometers (such as LIGO) it is the Michelson
topology and in DFIs proposed in Ref.
\cite{2006_interferometers_DNF_GW_detection} it is the Mach-Zehnder
(MZ) topology.

The analysis performed by S. Kawamura \textit{et al.} in Ref.
\cite{2006_interferometers_DNF_GW_detection} showed, however, that
though it is possible to eliminate all the information about
displacement and laser noises from the data, the sensitivity to GWs
at low frequencies turns out to be limited by the
$(\omega_{\textrm{gw}}L/c)^2$-factor for 3D (space-based)
configurations and $(\omega_{\textrm{gw}}L/c)^3$-factor for 2D
(ground-based) configurations. In the latter case this means the
cancelation of all the terms of the order of
$h(L/\lambda_{\textrm{gw}})^n$, $n=0,1,2$. For the signals around
$\omega_{\textrm{gw}}/2\pi\approx 100$ Hz and $L\approx 4$ km, the
DFI sensitivity of the ground-based detector is $\sim 10^{6}$ times
worse than the one of the conventional Michelson interferometer
(i.e. a single round-trip detector). The proposed MZ-based
configurations could be modified with power- and signal-recycling
mirrors, artificial time-delay devices \cite{2007_time_delay}, but
nevertheless, the potentially achievable sensitivity remains
incomparable with conventional non-DFI detectors. However, it is
worth noting that the basic features of DFI concept has been
recently demonstrated experimentally
\cite{2007_DFI_exp_1,2007_DFI_exp_2}.

In this paper we continue investigation of the noise cancelation
issue in large-scale interferometric experiments and present the
result of intensive discussions inside the GW community
\cite{2007_DPFP_LSC_1,2007_DPFP_LSC_2}. Namely, we propose a simple
toy model of the GW detector \textit{partially} free from
displacement noise of the test masses with strong enough GW
response. It should be stressed that our model is now widely
discussed with regard to its implementation in the 3rd generation GW
detectors and may soon be realized in a prototype lab experiment
\cite{2008_Virt_GW}. The basic element of our model is a single
detuned Fabry-Perot (FP) cavity pumped through both of its movable,
partially transparent mirrors; lasers and detectors are assumed to
be located on auxiliary (also movable) platforms. Pump waves in
different input ports are assumed to be orthogonally polarized in
order the corresponding output waves to be separately detectable and
to exclude nonlinear coupling of the corresponding intracavity
waves. By properly combining the signals of all four output ports of
the cavity (a pair of reflection and transmission ports for each of
the pumps) an experimentalist can remove the information about the
fluctuations of the mirrors coordinates from the data. Below we call
the proposed scheme a double-pumped Fabry-Perot (DPFP) cavity. In
this paper we do not consider the problem of optical laser noise
cancelation in full detail and thus ``displacement noise'' refers
only to mechanical motions of the test masses further. However,
after detailed analysis of a single DPFP cavity we propose and
consider qualitatively one of the possible DPFP-based balanced
optical setups, namely a LIGO-type topology of Michelson
interferometer with two DPFP cavities in its arms.

The isolation of the GW signal from displacement noise in a DPFP
cavity is achieved in a different manner as compared to MZ-based
interferometers. The basic idea is that when a detuned FP cavity is
pumped through one of the mirrors (mirror $a$ for definitness), the
reflected and transmitted waves respond differently to the motion of
mirrors $a$ and $b$. The physical reason for this is that the
reflected wave, in contrast to the transmitted one, includes the
component due to the prompt reflection from mirror $a$. This
component measures only the position of mirror $a$ but not the
position of mirror $b$. By properly combining both the response
signals one can eliminate the information about the fluctuating
coordinate of mirror $a$ completely, leaving only the part of the
signal containing the displacement noise of mirror $b$ plus its
displacement due to GW (assuming we work in the local Lorentz frame
of mirror $a$). By pumping the cavity through mirror $b$ and
performing the similar operations, one can eliminate the information
about displacement noise of mirror $b$. Ultimately, the proper
linear combination of all four output signals cancels displacement
noise of both the mirrors leaving a non-vanishing GW signal.

In the resonant regime both the response signals (corresponding to
one of the pumps) carry identical information about the mirrors
coordinates and thus cannot be combined to cancel their
fluctuations. This happens because the prompt reflection does not
occur for the resonant pump.

Note that the LL-effect of GW direct coupling to light plays no role
in this noise-cancelation scheme: the notion of the GW in our
analysis can be approximated with the corresponding tidal
force-field. This means that the leading order of the DFI signal we
obtain will be $h(L/\lambda_{\textrm{gw}})^0$.

The ``payment'' for isolation of the GW signal from displacement
noise in our case is the loss of the optical resonant gain of the
order of $c/(\gamma L)$, where $\gamma$ is the cavity
half-bandwidth. In conventional interferometers this resonant factor
describes the accumulation of the low-frequency GW signal by the
light wave circulating in a FP cavity. The DFI response signal of a
DPFP cavity becomes limited with the factor of the order of unity as
compared to the limiting factor $(\omega_{\textrm{gw}}L/c)^3\sim
6\times 10^{-7}$ of the double Mach-Zehnder configuration
\cite{2006_interferometers_DNF_GW_detection} for $L\approx 4$ km and
$\omega_{\textrm{gw}}/2\pi\approx 100$ Hz. This difference between
the MZ-based topologies and the DPFP topology arises due to the
different mechanisms of noise cancelation: the former utilizes the
LL-effect of direct interaction between the GW and light, while the
latter utilizes the asymmetry between the output signals of detuned
cavity.

However, the most dramatic consequence which the loss of the
resonant gain results in is that the displacement noise of the
auxiliary platforms (where lasers and detectors are mounted) becomes
comparable to the DFI response. The reason for this is the
relativity principle itself: only relative measurements of the test
masses positions and velocities are allowed; in our case we are able
to measure the positions of cavity mirrors only with respect to the
mentioned auxiliary platforms. It is natural then that the precision
of the coordinate measurements is limited with the noises of
reference test masses (see Sec. \ref{sec_mechanism} below). Remind
also that in conventional non-DFI (LIGO) topology these noises are
negligible since they are suppressed finesse times as compared to
the GW signal (and displacement noise of the mirrors). The
incomplete cancelation of displacement noise is the major
(fundamental) limitation of our model. To increase its SNR in
practice one will need to install lasers and detectors on heavy
platforms (to suppress displacement noise due to external forces)
cooled down to cryogenic temperatures (to suppress internal thermal
noise).

Note that the non-resonant regime implies the rise of the
electromagnetic ponderomotive force (and corresponding optical
rigidity) acting on the mirrors of a FP cavity
\cite{1967_EM_rigidity,1997_optical_bars,1999_rigidity_QM,
2000_rakh_phd_thesis,2001_FD_rigidity,2001_quantum_noise,
2001_SR_optical_springs,2005_optical_rigidity_SR,2006_sub_SQL}.
However, in this paper we do not take into account the effects of
radiation pressure. In particular, optical rigidity vanishes if pump
waves in different input ports have detunings with equal absolute
values but opposite signs.

This paper is split into two logical parts for convenience. First,
in Sec. \ref{sec_mechanism} using a simple mathematical model of the
cavity we illustrate the basic physics underlying the proposed
method of noise cancelation. Next, in Secs. \ref{sec_space_time_gen}
--- \ref{sec_DPFP} we perform strict calculations. Finally, in Sec.
\ref{sec_discussion} we discuss some issues associated with
displacement noise cancelation, consider Michelson/DPFP balanced
optical setup for laser optical noise cancelation and briefly
outline further prospects.

\section{Basic physical mechanism of noise cancelation}\label{sec_mechanism}
Before analyzing our scheme in full detail we consider the simplest
(Newtonian) model of a FP cavity to demonstrate the basic physics
underlying the mechanism of noise cancelation.

Let us derive the response signals of a FP cavity from the intuitive
reasonings assuming that the cavity is short enough (we neglect time
delays) and the GW can be treated as a classical force acting on the
test masses.
\begin{figure}[h]
\begin{center}
\includegraphics[scale=0.7]{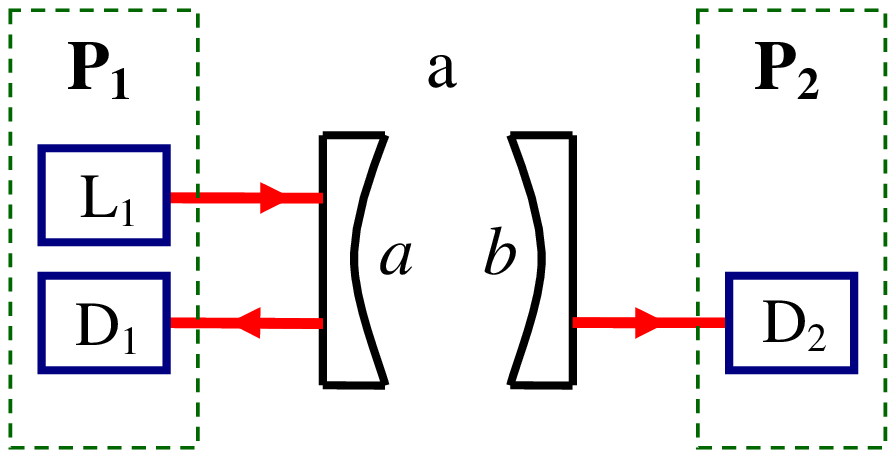}
\includegraphics[scale=0.7]{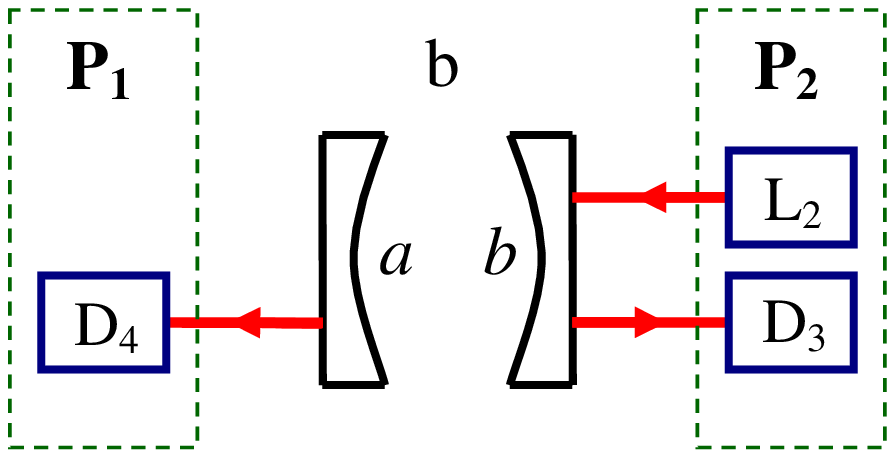}
\caption{A simple noise-cancelation setup. a. Fabry-Perot cavity
assembled of two movable, partially transparent, mirrors $a$ and $b$
is pumped by laser $\textrm{L}_1$ through mirror $a$. Detectors
$\textrm{D}_1$ and $\textrm{D}_2$ measure the phases of reflected
and transmitted waves correspondingly. b. The same cavity is pumped
by laser $\textrm{L}_2$ through mirror $b$. Detectors $\textrm{D}_3$
and $\textrm{D}_4$ measure the phases of reflected and transmitted
waves correspondingly.} \label{pic_simple_FP}
\end{center}
\end{figure}
Consider a system illustrated in Fig. \ref{pic_simple_FP}a: FP
cavity assembled of two movable, partially transparent, mirrors $a$
and $b$ is pumped by laser $\textrm{L}_1$ through mirror $a$.
Detectors $\textrm{D}_1$ and $\textrm{D}_2$ measure the phases of
reflected and transmitted waves correspondingly. For simplicity we
assume that laser $\textrm{L}_1$ and detector $\textrm{D}_1$ are
rigidly mounted on platform $\textrm{P}_1$ and detector
$\textrm{D}_2$ is rigidly mounted on platform $\textrm{P}_2$. It is
evident that the wave circulating inside the cavity measures the
relative displacement of mirrors $a$ and $b$ plus GW displacement:
$\xi_{\textrm{gw}}+\xi_b-\xi_a$. Transmitted signal
$a^\textrm{t}_{\textrm{out}}$ can be measured in such a way that it
will be directly proportional to this quantity:
\begin{subequations}
\begin{equation}
    a^\textrm{t}_{\textrm{out}}=q_1(\xi_{\textrm{gw}}+\xi_b-\xi_a).
    \label{simple_trans_sig}
\end{equation}
Here $q_1$ describes the resonant gain (amplification of the phase
shift) of the cavity. The reflected signal
$a^\textrm{r}_{\textrm{out}}$ is somewhat different: it also
includes the component due to the prompt reflection of the pump wave
from the input mirror. For instance, if the cavity is pumped through
mirror $a$ then this component is proportional to
$\xi_a-\xi_{\textrm{P}_1}$. The reflected signal is then
\begin{equation}
    a^\textrm{r}_{\textrm{out}}=p(\xi_a-\xi_{\textrm{P}_1})+
    q_2(\xi_{\textrm{gw}}+\xi_b-\xi_a).
    \label{simple_refl_sig}
\end{equation}
\end{subequations}
Here $q_2$ also describes the resonant gain (multiple reflections
inside the cavity), while $p$ is the quantity of the order of unity
since it describes a single reflection from the input mirror.
Equations (\ref{simple_trans_sig}) and (\ref{simple_refl_sig}) tell
us that we are unable to measure absolute values of $\xi_a$ and
$\xi_b$, only relative measurements, e.g. with respect to platform
$\textrm{P}_1$, are allowed.

Now consider the situation illustrated in Fig. \ref{pic_simple_FP}b:
the same cavity is pumped by laser $\textrm{L}_2$ through mirror $b$
with the wave polarized normally to the wave emitted by laser
$\textrm{L}_1$. Detectors $\textrm{D}_3$ and $\textrm{D}_4$ measure
the phases of reflected and transmitted waves correspondingly. Again
we assume that laser $\textrm{L}_2$ and detector $\textrm{D}_3$ are
rigidly mounted on platform $\textrm{P}_2$ and detector
$\textrm{D}_4$ is rigidly mounted on platform $\textrm{P}_1$. The
second pair of response signals can be derived in full similarity.
Let us consider the simplest case of equal pumps (equal amplitudes
and detunings). Then due to the symmetry of the system and plane GW
wavefront the second pair of responses can be written as:
\begin{subequations}
\begin{align}
    b^\textrm{t}_{\textrm{out}}&=q_1(\xi_{\textrm{gw}}+\xi_b-\xi_a),
    \label{simple_trans_sig_2}\\
    b^\textrm{r}_{\textrm{out}}&=p(\xi_{\textrm{P}_2}-\xi_b)+
    q_2(\xi_{\textrm{gw}}+\xi_b-\xi_a)\label{simple_refl_sig_2}.
\end{align}
\end{subequations}
Here displacements of the mirrors are measured with respect to
platform $\textrm{P}_2$.

Now constructing the following linear combination of the responses
\begin{equation*}
    s=a^\textrm{r}_{\textrm{out}}+\frac{p-q_2}{q_1}\,a^\textrm{t}_{\textrm{out}}+
    b^\textrm{r}_{\textrm{out}}-\frac{q_2}{q_1}\,b^\textrm{t}_{\textrm{out}},
\end{equation*}
we are able to cancel displacement noise of both mirrors:
\begin{equation}
    s=p(\xi_{\textrm{gw}}+\xi_{\textrm{P}_2}-\xi_{\textrm{P}_1}).
    \label{simple_DFI}
\end{equation}
Note that displacement noise of the platforms cannot be eliminated.
This is the direct consequence of the \textit{relativity principle}
which states that no absolute coordinate or velocity measurements
are allowed: one can measure the coordinates of the mirrors only
with respect to the positions of reference test masses, platforms
$\textrm{P}_1$ and $\textrm{P}_2$ in our case. Therefore, it is
natural that displacement noise of the reference masses imposes the
sensitivity limit of the coordinate measurements.

According to formula (\ref{simple_DFI}), noise cancelation in a DPFP
cavity is possible due to the effect of prompt reflection from the
input mirror which is described by the $p$-multiplier. The obtained
DFI response is similar to the response of a simple single-pass GW
detector: an observer sends the light wave to the reflective mirror
and receives it back measuring the phase shift. The
noise-cancelation algorithm that we perform for a DPFP cavity in
some sense can be interpreted as removal of the cavity ``by hands''.
Evidently, this results in the loss of the optical resonant gain:
signal $s$ in formula (\ref{simple_DFI}) includes neither $q_1$ nor
$q_2$.

Two special cases when noise cancelation is impossible can be
immediately ``predicted'' from Eqs. (\ref{simple_trans_sig} ---
\ref{simple_refl_sig_2}): (i) $p=0$, meaning that the prompt
reflection does not occur (this takes place for the resonant pump,
see below) and (ii) $\xi_a=\xi_{\textrm{P}_1}$ and simultaneously
$\xi_b=\xi_{\textrm{P}_2}$, meaning that the mirrors are rigidly
attached to the platforms.

It is evident now that the relativity principle and the notion of
the reference frame play significant roles in our analysis. The
simplest model of the cavity presented above has been considered in
the laboratory (globally inertial) reference frame. However, such a
consideration is not free from certain drawbacks. In particular,
formula (\ref{simple_trans_sig}) for the transmitted signal cannot
be derived from the simple assumption that the phase of transmitted
wave (after emission at $\textrm{P}_1$) is measured at platform
$\textrm{P}_2$ with respect to the local clocks (i.e. clocks located
at $\textrm{P}_2$). To justify formula (\ref{simple_trans_sig})
certain manipulations with the reference wave, produced by laser
$\textrm{L}_1$, need to be performed. However, if one wishes to
calculate \textit{in the laboratory frame} the phase of transmitted
signal at $\textrm{P}_2$ with respect to the local clocks (i.e. the
reference oscillation produced at $\textrm{P}_2$), he will
inevitably run into a ``forward-trip paradox'' described and
resolved in Ref. \cite{accel_observ}: if platforms $\textrm{P}_1$
and $\textrm{P}_2$ move as a single body (i.e.
$\xi_{\textrm{P}_1}=\xi_{\textrm{P}_2}$) and cavity mirrors are
either absent or attached to the platforms, then the phase shift
carried by the transmitted wave will contain information about the
velocity of the whole system with respect to some ``absolute
space'', that is forbidden by the relativity principle. To avoid the
paradox one should perform the calculations in the proper reference
frame of detector which is non-inertial in general due to the action
of external fluctuative forces. The accelerated frame necessarily
implies the use of general relativity (GR). Therefore, to obtain
strict and consistent description of the cavity in general case we
need to complete fully general relativistic calculations.

Another reason to implement GR is the notion of the GW itself which
is a purely GR effect. Even though in this paper one may reduce the
action of the GW to the effective (Newtonian) tidal force-field,
further development of the DPFP idea suggests that displacement
noise of the auxiliary platforms may be eliminated but at the cost
of GW response reduction, leading to the
$(L/\lambda_{\textrm{gw}})^n$ limiting factors which have a purely
GR nature.

Furthermore, it is widely known that the fundamental limit of the
sensitivity of optical interferometers is imposed by the vacuum
photon shot noise: it will be the only limiting factor left when
other noises are canceled or suppressed. Therefore, in order to
analyze the ultimate sensitivity of our GW detector we need to
quantize the electromagnetic wave circulating inside the cavity
since vacuum noise cannot be obtained in the framework of classical
electrodynamics.

Summing up, one may conclude that the most general and strict
problem definition would be the boundary problem for quantized
electromagnetic wave in the space-time of accelerated observer in
the field of the GW. Therefore, in Sec. \ref{sec_space-time} we
define the mentioned space-time, then in Sec. \ref{sec_quant_wave}
we remind the formalism of the optical wave quantization and finally
in Sec. \ref{sec_FP_cavity} we set and solve the corresponding
boundary problem for a FP cavity.

\section{Space-time of accelerated observer in the gravitational
wave field}\label{sec_space_time_gen}
\subsection{Motion of the test masses}\label{sec_space-time}
In the Earth-bound GW observatories all the test masses including
lasers and detectors undergo fluctuative motions. Since it is the
detector that produces an experimentally observable quantity one
should consider the operation of an interferometer in its proper
reference frame, which is non-inertial in general. For this purpose
we first introduce the space-time associated with an observer having
non-geodesic 3-acceleration
$\ddot{\xi}_i(t)=\{\ddot{\xi}_x(t),\ddot{\xi}_y(t),\ddot{\xi}_z(t)\}$
and falling in the GW field $h=h(t-z/c)$. We assume the latter to be
weak, plane, '+'-polarized and propagating along the $z$-axis. The
case of generic GW polarization and direction of propagation does
not introduce any significant changes (in the context of this work)
to our further analysis. Therefore, in the proper reference frame of
such an observer space-time metric takes the following form
\cite{2003_clas_phys_acc_obs,1978_inert_grav_eff,1994_fermi,accel_observ,
2003_clas_phys_GW,2005_local_observ,2007_GW_FP_LL}:
\begin{align}
    ds^2=&-(c\,dt)^2\left[1+\frac{2}{c^2}\,\ddot{\xi}_i(t)x^i\right]+
    dx^2+dy^2+dz^2\nonumber\\
    &+\,\frac{1}{2}\,\frac{x^2-y^2}{c^2}\,\ddot{h}(t-z/c)\,
    (c\,dt-dz)^2.
    \label{metric_tensor}
\end{align}
Latin indices run over 1, 2, 3. In this paper we consider only
one-dimensional motion of the test masses, thus without the loss of
generality we may assume $y=z=0$ and denote $\xi_x(t)\equiv\xi(t)$.
In practice fluctuative forces acting on the test masses are very
weak, as the GW itself, thus it is natural to require that for all
reasonable $x$ and $t$ conditions $|2\ddot{\xi}x/c^2|\ll1$ and
$|h|\ll 1$ are fulfilled so we can use the methods of linearized
theory.

Metric (\ref{metric_tensor}) has two special cases.
\begin{enumerate}
\item $\ddot{\xi}(t)=0$ and the proper reference frame coincides with the local
Lorentz frame (also called the LL gauge in literature) of the
observer freely falling in the GW field. It is worth noting that the
LL gauge is free from the requirement of the distance $L$ between
the test masses to be much smaller than the gravitational wavelength
$\lambda_{\textrm{gw}}$ \cite{2005_local_observ,2007_GW_FP_LL}.
Corresponding approximation $L\ll\lambda_{\textrm{gw}}$ will be
called below the long-wave approximation.
\item $h(t)=0$ and the proper reference frame is simply a
non-inertial frame in Newtonian sense. Note that the curvature of
space-time with metric (\ref{metric_tensor}) equals to zero under
this condition, since it can be made globally flat with the
coordinate transformation that brings us from the non-inertial frame
to the inertial one.
\end{enumerate}

Remind, that due to the relativity principle an observer is unable
to measure his non-geodesic displacement $\xi(t)$ absolutely, in
contrast to the corresponding acceleration $\ddot{\xi}(t)$. To avoid
the ambiguity associated with the choice of initial conditions
$\xi(0)$ and $\dot{\xi}(0)$ we assume below that $\xi(t)$ is
measured in such a globally inertial (laboratory) reference frame in
the absence of the GW, that $\ddot{\xi}(t)=0$ results in $\xi(t)=0$.

The solution to geodesic equation corresponding to metric
(\ref{metric_tensor}) can be found in Ref. \cite{accel_observ}. If
the $j$th test mass and an observer are separated by a distance $L$
on the average (the 0th order solution) then the test mass
displacement relative to an observer (the 1st order solution) equals
to $X_j(t)=\frac{1}{2}Lh(t)-\xi(t)$. If, in addition, the test mass
is subjected to some non-GW forces and undergoes corresponding
fluctuative displacement $\xi_j(t)$ (measured in the globally
inertial reference frame) then
\begin{equation}
    X_j(t)=\frac{1}{2}\,Lh(t)+\xi_j(t)-\xi(t).
    \label{law_of_motion}
\end{equation}
Below we assume that for any test mass both its displacements
$X_j(t)$ and $\xi_j(t)$ obey the relation $|X_j|,\ |\xi_j|\ll L$. We
will also widely use the spectral domain where
\begin{equation*}
    \begin{bmatrix}
        X_j(t)\\
        \xi_j(t)\\
    \end{bmatrix}=
        \int_{-\infty}^{+\infty}
    \begin{bmatrix}
        X_j(\Omega)\\
        \xi_j(\Omega)\\
    \end{bmatrix}e^{-i\Omega t}\,\frac{d\Omega}{2\pi}.
\end{equation*}

The introduced proper reference frame is the best suited for
analysis of the GW detectors with the test masses undergoing
non-geodesic motion, in contrast to the transverse-traceless (TT)
gauge, where such an analysis should be additionally validated. In
addition, proper reference frame is the natural frame used by
Newtonian experimentalists performing measurements in the laboratory
and recording the obtained data from detectors.

\subsection{Quantized electromagnetic wave interacting with
the weak gravitational wave in a non-inertial
frame}\label{sec_quant_wave} In the interferometric experiments an
observer studies the motion of the test masses by sending and
receiving the reflected light waves. Thus it is necessary to take
into account the effects imposed by the GW and acceleration fields
on the optical field for a complete description of an
interferometer. Here we briefly remind the formalism used to
describe the quantized electromagnetic wave (EMW) propagating in the
space-time with metric (\ref{metric_tensor}).

First, we start from the simplest case of Minkowski space-time. It
is convenient to represent the electric field operator of the EMW as
a sum of (i) the ``strong'' (classical) plane monochromatic wave
(which approximates the light beam with cross-section $S$) with
amplitude $A_0$ and frequency $\omega_0$ and (ii) the ``weak'' wave
describing quantum fluctuations of the electromagnetic field (see
Appendix \ref{app_quant_emw}):
\begin{subequations}
\begin{align*}
    A(x,t)&=\sqrt{\frac{2\pi\hbar\omega_0}{Sc}}\,
    \Bigl[A_0+a(x,t)\Bigr]e^{-i(\omega_0t\mp k_0x)}+{\textrm{h.c.}},\\
    a(x,t)&=\int_{-\infty}^{+\infty}
    a(\omega_0+\Omega)e^{-i\Omega\left(t\mp x/c\right)}\,\frac{d\Omega}{2\pi},
\end{align*}
\end{subequations}
with amplitude $a(\omega_0+\Omega)$ (Heisenberg operator to be
strict) obeying the commutation relations:
\begin{align*}
    \bigl[a(\omega_0+\Omega),a(\omega_0+\Omega')\bigr]&=0,\\
    \bigl[a(\omega_0+\Omega),a^\dag(\omega_0+\Omega')\bigr]&=2\pi\delta(\Omega-\Omega').
\end{align*}
This notation for quantum fluctuations $a(x,t)$ will be the most
suitable for us since it coincides exactly with the
Fourier-representation of the classical fields. For briefness
throughout the paper we omit the
$\sqrt{2\pi\hbar\omega_0/Sc}$-multiplier and notation ``h.c.'' We
call $A(x,t)$ the vacuum-state wave if $A_0=0$.

Electromagnetic wave propagating in space-time with metric
(\ref{metric_tensor}) directly couples to the GW and acceleration
fields. We will study only the 1st order (in $h$ and $\xi$) coupling
effects and neglect the GW and acceleration interaction with the
optical noise. In other words, both the GW and acceleration fields
are assumed to be coupled only to the ``strong'' (classical) wave
\cite{2007_GW_FP_LL,accel_observ}:
\begin{align}
    A(x,t)=
    &\Bigl[A_0+A_0g_{\pm}(x,t)+A_0w_{\pm}(x,t)+a(x,t)\Bigr]\nonumber\\
    &\times e^{-i(\omega_0t\mp k_0x)},\label{EMW}
\end{align}
where
\begin{align*}
    g_{\pm}(x,t)&=\int_{-\infty}^{+\infty}g_\pm(x,\omega_0+\Omega)e^{-i\Omega t}\,
    \frac{d\Omega}{2\pi},\\
    g_\pm(x,\omega_0+\Omega)&=h(\Omega)
    \biggl[\frac{1}{4}\,\omega_0\Omega\,\frac{x^2}{c^2}\mp i\,\frac{1}{2}\,k_0x\nonumber\\
    &\qquad\qquad+\frac{1}{2}\,\frac{\omega_0}{\Omega}\,
    \left(e^{\pm i\Omega x/c}-1\right)\biggr],
\end{align*}
and
\begin{align*}
    w_{\pm}(x,t)&=\int_{-\infty}^{+\infty}w_\pm(x,\omega_0+\Omega)e^{-i\Omega t}\,
    \frac{d\Omega}{2\pi},\\
    w_\pm(x,\Omega+\omega_0)&=-k_0\xi(\Omega)
    \biggl[\frac{\Omega}{c}\,x\pm i\Bigl(e^{\pm i\Omega x/c}-1\Bigr)\biggr].
\end{align*}
Both $g_\pm(x,t)$ and $w_\pm(x,t)$ describe the distributed effects:
$g_\pm$ is responsible for the direct coupling between the GW and
the EMW and $w_\pm$ describe the redshift imposed on the EMW by the
non-inertiality of the reference frame. Both terms are accurate up
to the order of $(\Omega/\omega_0)^0$, vanish at $x=0$ and in
long-wave (or low-frequency) approximation have the $O[(\Omega
x/c)^2]$ asymptotics. It is also straightforward to verify that both
$g_\pm(x,t)$ and $w_\pm(w,t)$ are the pure imaginary values;
sometimes it will be convenient to use the following approximate
formulas:
\begin{subequations}
\begin{align}
    1+g_\pm(x,t)&=1+i\mathfrak{I}\bigl[g_\pm(x,t)\bigr]
    \approx e^{i\mathfrak{I}\bigl[g_\pm(x,t)\bigr]},\label{gw+emw_imag}\\
    1+w_\pm(x,t)&=1+i\mathfrak{I}\bigl[w_\pm(x,t)\bigr]
    \approx e^{i\mathfrak{I}\bigl[w_\pm(x,t)\bigr]}.\label{acc+emw_imag}
\end{align}
\end{subequations}

\section{Response of a Fabry-Perot cavity to a plane gravitational wave}\label{sec_FP_cavity}
\subsection{Input, circulating and output waves}
\begin{figure}[h]
\begin{center}
\includegraphics[scale=0.7]{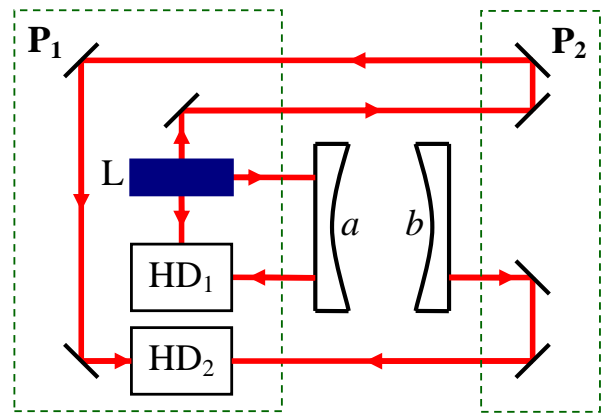}
\caption{Emission-detection scheme. Pump wave is radiated by laser L
and reflected wave is detected with the homodyne detector
$\textrm{HD}_1$. Transmitted wave is redirected towards platform
$\textrm{P}_1$ and is detected with the homodyne detector
$\textrm{HD}_2$. Laser L and both the homodyne detectors are assumed
to be rigidly mounted on platform $\textrm{P}_1$. Mirrors which
redirect the transmitted wave (and the reference wave) towards
detector $\textrm{HD}_2$ are assumed to be rigidly mounted on
platform $\textrm{P}_2$.} \label{pic_emission_detection}
\end{center}
\end{figure}

Let us consider the operation of the optical scheme, illustrated in
Fig. \ref{pic_emission_detection}, which consists of platforms
$\textrm{P}_{1,2}$ and a FP cavity assembled of two movable mirrors
$a$ and $b$, both lossless and having the amplitude transmission
coefficient $T$, $|T|\ll1$. We put distance between the mirrors in
the absence of the gravitational wave and optical radiation to be
equal to $L$. Without the loss of generality we assume the cavity to
be lying in the plane $z=0$ along one of the GW principal axes,
coinciding with the $x$-axis.

Laser L and the homodyne detectors $\textrm{HD}_{1,2}$ are assumed
to be \textit{rigidly} mounted on platform $\textrm{P}_1$. In other
words, we assume that all the elements on the platform do not move
with respect to each other. Similarly, the auxiliary mirrors which
redirect the transmitted wave (see below) are rigidly mounted on
platform $\textrm{P}_2$. We introduce these requirements into our
toy model in order not to deal with the inessential relative motions
of the optical scheme elements. However, in practice these motions
will result in some additional displacement noise.

In this section we will work in the proper reference frame of (the
center of mass of) platform $\textrm{P}_1$ at which the origin of
the coordinate system is set: $x_{\textrm{P}_1}(t)=0$. Then the
coordinates (their operators to be strict) of the mirrors are
$x_a(t)=l_1+X_a(t)\approx X_a(t)$ and $x_b(t)=L+l_1+X_b(t)\approx
L+X_b(t)$, where $l_1\ll L$ is the negligible distance between the
center of mass of platform $\textrm{P}_1$ and mirror $a$. The
coordinate of (the center of mass of) platform $\textrm{P}_2$ is
$x_{\textrm{P}_2}(t)=L+l_1+l_2+X_{\textrm{P}_2}(t)\approx
L+X_{\textrm{P}_2}(t)$, where $l_2\ll L$ is the negligible distance
between the center of mass of platform $\textrm{P}_2$ and mirror
$b$. Remind that $X_{a,b,\textrm{P}_2}(t)$ are the displacements
with respect to non-inertial reference frame of platform
$\textrm{P}_1$ and obey the relation $|X_{a,b,\textrm{P}_2}|\ll L$.

\begin{figure}[h]
\begin{center}
\includegraphics[scale=0.57]{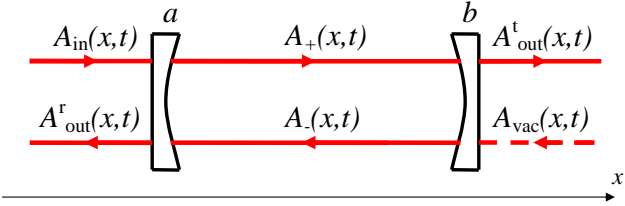}
\caption{Fabry-Perot cavity assembled of two movable mirrors $a$ and
$b$. Cavity is pumped through mirror $a$ with the input wave
$A_{\textrm{in}}(x,t)$ and through mirror $b$ with the vacuum-state
wave $A_{\textrm{vac}}(x,t)$. Optical field inside the cavity is
represented as a sum of the wave $A_{+}(x,t)$, running in the
positive direction of the $x$-axis, and the wave $A_{-}(x,t)$,
running in the opposite direction. The reflection-output signal is
$A_{\textrm{out}}^{\textrm{r}}(x,t)$ and transmission-output signal
is $A_{\textrm{out}}^{\textrm{t}}(x,t)$.}
\label{pic_Fabry-Perot_cavity}
\end{center}
\end{figure}
Let the cavity be pumped by laser L through mirror $a$ with the
input wave (see Fig. \ref{pic_Fabry-Perot_cavity})
\begin{align}
    A_{\textrm{in}}(x,t)=
    &A_{\textrm{in}0}\Bigl[1+g_+(x,t)+w_+(x,t)\Bigr]e^{-i(\omega_1 t-k_1x)}\nonumber\\
    &+a_{\textrm{in}}(x,t)e^{-i(\omega_1 t-k_1x)},\label{input_wave}
\end{align}
and with the vacuum-state wave through mirror $b$:
\begin{equation}
    A_{\textrm{vac}}(x,t)=a_{\textrm{vac}}(x,t)e^{-i\bigl[\omega_1 t+k_1(x-L)\bigr]},
    \label{vacuum}
\end{equation}
Here $a_{\textrm{in}}(x,t)$ is the ``weak'' field describing laser
noise of the pump wave and $a_{\textrm{vac}}(x,t)$ is the ``weak''
field describing vacuum noise in the opposite input port. Remind,
that both the laser and mirror $a$ are located at $x\approx0$, where
$g(0,t)=w(0,t)=0$, thus input wave does not acquire distributed
phase shift when it reaches mirror $a$.

It is convenient to represent the optical field inside the cavity as
a sum of two waves, $A_{+}(x,t)$ and $A_{-}(x,t)$, running in the
opposite directions:
\begin{align}
    A_{\pm}(x,t)=&A_{\pm0}\Bigl[1+g_\pm(x,t)+w_\pm(x,t)\Bigr]
    e^{-i(\omega_1 t\mp k_1x)}\nonumber\\
    &+a_{\pm}(x,t)e^{-i(\omega_1 t\mp k_1x)}.\label{inside_wave}
\end{align}
Here $a_\pm(x,t)$ describes the phase shift accumulated by the light
wave while circulating inside the cavity.

Output wave reflected from the cavity is:
\begin{align}
    A^{\textrm{r}}_{\textrm{out}}(x,t)
    =&A^{\textrm{r}}_{\textrm{out}0}\Bigl[1+g_-(x,t)+w_-(x,t)\Bigr]
    e^{-i(\omega_1t+k_1x)}\nonumber\\
    &+a^{\textrm{r}}_{\textrm{out}}(x,t)e^{-i(\omega_1 t+k_1x)},
    \label{reflected_wave}
\end{align}
Quadrature components (see Appendix \ref{app_quant_emw}) of this
wave are assumed to be measured with the homodyne detector
$\textrm{HD}_1$ (see Fig. \ref{pic_emission_detection}). The
reference oscillation is produced by laser L.

Output wave transmitted through the cavity
\begin{align}
    A^{\textrm{t}}_{\textrm{out}}(x,t)
    =&A^{\textrm{t}}_{\textrm{out}0}\Bigl[1+g_+(x,t)+w_+(x,t)\Bigr]
    e^{-i\bigl[\omega_1 t-k_1(x-L)\bigr]}\nonumber\\
    &+a^{\textrm{t}}_{\textrm{out}}(x,t)e^{-i\bigl[\omega_1t-k_1(x-L)\bigr]},
    \label{transmitted_wave}
\end{align}
is redirected towards platform $\textrm{P}_1$ by the small auxiliary
mirrors mounted on platform $\textrm{P}_2$. Quadratures of the
transmitted wave are measured with the homodyne detector
$\textrm{HD}_2$ (see Fig. \ref{pic_emission_detection}). The
reference oscillation is produced by laser L which commits a single
round trip along the $\textrm{P}_1-\textrm{P}_2-\textrm{P}_1$ path
(see below).

It should be mentioned that since both the reflected and transmitted
waves commit round-trips and are detected at location of the source,
one may perform all the calculations in the TT gauge (see Ref.
\cite{accel_observ}). However, for the sake of generality we work in
the proper reference frame of detector.

Note that the complex amplitudes
$a^{\textrm{r,t}}_{\textrm{out}}(x,t)$ are the unknown function of
their arguments and are obtained as the solutions of the
corresponding boundary problem for a FP cavity (see below).
Obviously, they should vanish in the limit $R\rightarrow 0$, i.e. in
the absence of the cavity, if
$a_{\textrm{in}}=a_{\textrm{vac}}\equiv 0$. Therefore, below we call
functions $a^{\textrm{r,t}}_{\textrm{out}}(x,t)$ or
$a^{\textrm{r,t}}_{\textrm{out}}(\omega_1+\Omega)$ the cavity
response (or output) signals, meaning that they describe the
influence of a FP cavity on the light propagation. The summand
proportional to $A_{\textrm{out}0}^{\textrm{r}}$ in formula
(\ref{reflected_wave}) and the one proportional to
$A_{\textrm{out}0}^{\textrm{t}}$ in (\ref{transmitted_wave}) thus
correspond to the ``no-cavity'' case and are unimportant for us. In
order to make our analysis more transparent we construct our
detection scheme in such a way that these terms become unmeasurable.

In the case of reflected wave both $g_-(x,t)$ and $w_-(x,t)$ vanish
at $x=0$ and the only measurable quantities left are the quadratures
of $a^{\textrm{r}}_{\textrm{out}}(x,t)$.

The case of transmitted wave is more complex. Note that the
$A_{\textrm{out}0}^{\textrm{t}}$-summand in formula
(\ref{transmitted_wave}) at point $x=x_{\textrm{P}_2}(t)$ describes
a \textit{single} forward trip of light along the cavity:
\begin{multline*}
    \Bigl[1+g_{+}(x_{\textrm{P}_2},t)+w_{+}(x_{\textrm{P}_2},t)\Bigr]
    e^{ik_1X_{\textrm{P}_2}(t)}\\
    \approx\exp\biggl\{ik_1X_{\textrm{P}_2}(t)+
    i\mathfrak{I}\Bigl[g_+(L,t)+w_+(L,t)\Bigr]\biggr\}.
\end{multline*}
Here we used formulas (\ref{gw+emw_imag}) and (\ref{acc+emw_imag}).
Remind also, that the transmitted wave is redirected towards
platform $\textrm{P}_1$ for detection and thus commits a backward
trip. Clearly, the whole round trip will result in phase shift
\begin{equation*}
    2k_1X_{\textrm{P}_2}(t)+\mathfrak{I}\Bigl[g_+(L,t)-g_-(L,t)+w_+(L,t)-w_-(L,t)\Bigr].
\end{equation*}
In order to make this phase shift unmeasurable we make the reference
wave, produced by laser L, to travel the same round trip before
returning to the homodyne detector $\textrm{HD}_2$. Ultimately, both
the additional phases of the transmitted wave and of the reference
oscillation are completely subtracted in the homodyne measurement.
Therefore, the only measurable quantities left in the transmitted
wave are the quadratures of $a^{\textrm{t}}_{\textrm{out}}(x,t)$.

It is worth noting that such a detection scheme (illustrated in Fig.
\ref{pic_emission_detection}) only serves a purpose of making the
\textit{theoretical} (rather general) analysis of our toy model more
transparent. Experimentalists may want to change it in the way to
simplify this or that specific \textit{experimental} setup; small
changes in formulas need to be introduced then, depending on it.

\subsection{Response signals of a Fabry-Perot cavity}
To obtain the response functions of a Fabry-Perot cavity we
substitute fields (\ref{input_wave} -- \ref{transmitted_wave}) into
the set of boundary conditions (conditions of the electric field
continuity along the surfaces of the mirrors)
\cite{1995_mirror_radiation,2007_GW_FP_LL}:
\begin{subequations}
\begin{align}
    A_+(x_a,t)&=TA_{\textrm{in}}(x_a,t)-RA_-(x_a,t),\label{bound_a_1}\\
    A^{\textrm{r}}_{\textrm{out}}(x_a,t)&=RA_{\textrm{in}}(x_a,t)+TA_-(x_a,t),
    \label{bound_a_2}\\
    A_-(x_b,t)&=TA_{\textrm{vac}}(x_b,t)-RA_+(x_b,t),\label{bound_b_1}\\
    A^{\textrm{t}}_{\textrm{out}}(x_b,t)&=RA_{\textrm{vac}}(x_b,t)+TA_+(x_b,t).
    \label{bound_b_2}
\end{align}
\end{subequations}
This set of equations is accurate up to the 0th order of
$\Omega/\omega_1$ since it does not take into account the
relativistic terms proportional to $\dot{X}_{a,b}/c$
\cite{2007_GW_FP_LL}. The solution of this set is obtained in
Appendix \ref{app_boundaries} using the method of successive
approximations. Since we do not consider the effect of parametric
excitation of the additional optical modes under the influence of
the GW \cite{2007_GW_FP_LL}, it will be convenient to introduce the
detuning $\delta_1=\omega_1-\pi n_0/\tau$, where $n_0$ is integer,
even (for simplicity) and fixed; $\tau=L/c$. Then the solution of
the 1st order takes the following form (all spectral arguments are
omitted):
\begin{widetext}
\begin{subequations}
\begin{align}
    a^{\textrm{r}}_{\textrm{out}}&=
    \frac{R-Re^{2i(\delta_1+\Omega)\tau}}{1-R^2e^{2i(\delta_1+\Omega)\tau}}\,
    a_{\textrm{in}}
    +\frac{T^2e^{i(\delta_1+\Omega)\tau}}{1-R^2e^{2i(\delta_1+\Omega)\tau}}\,
    a_{\textrm{vac}}-
    \frac{RT^2A_{\textrm{in}0}e^{2i\delta_1\tau}}{1-R^2e^{2i\delta_1\tau}}\,i\,
    \frac{2k_1(X_be^{i\Omega\tau}-\sigma_1 X_a)+
    \delta\Psi_{\textrm{emw}}}{1-R^2e^{2i(\delta_1+\Omega)\tau}},
    \label{reflected_signal}\\
    a^{\textrm{t}}_{\textrm{out}}&=
    \frac{T^2e^{i(\delta_1+\Omega)\tau}}{1-R^2e^{2i(\delta_1+\Omega)\tau}}\,
    a_{\textrm{in}}
    +\frac{R-Re^{2i(\delta_1+\Omega)\tau}}{1-R^2e^{2i(\delta_1+\Omega)\tau}}\,
    a_{\textrm{vac}}
    +\frac{R^2T^2A_{\textrm{in}0}e^{3i\delta_1\tau}}{1-R^2e^{2i\delta_1\tau}}\,i\,
    \frac{2k_1(X_be^{i\Omega\tau}-X_a)+
    \delta\Psi_{\textrm{emw}}}{1-R^2e^{2i(\delta_1+\Omega)\tau}}\,
    e^{i\Omega\tau}.\label{transmitted_signal}
\end{align}
\end{subequations}
\end{widetext}
Here phase shift $\delta\Psi_{\textrm{emw}}=
\delta\Psi_{\textrm{gw+emw}}+\delta\Psi_{\textrm{acc+emw}}$,
calculated in the approximation $\Omega/\omega_1\ll1$, describes the
direct coupling of the optical wave to the GW and acceleration
fields:
\begin{subequations}
\begin{align}
    \delta\Psi_{\textrm{gw+emw}}(\Omega)&=
    -k_1Lh(\Omega)\left(1-\frac{\sin\Omega\tau}{\Omega\tau}\right)e^{i\Omega\tau},
    \label{gw+emw_phase}\\
    \delta\Psi_{\textrm{acc+emw}}(\Omega)&=
    -k_1\xi_{\textrm{P}_1}(\Omega)\Bigl(1-2e^{i\Omega\tau}+e^{2i\Omega\tau}\Bigr).
    \label{acc+emw_phase}
\end{align}
\end{subequations}
Remind that $\xi_{\textrm{P}_1}(\Omega)$ is the fluctuative
displacement of platform $\textrm{P}_1$ measured in the laboratory
frame. Factor
\begin{align*}
    \sigma_1(\Omega)&=e^{-2i\delta_1\tau}/T^2\\
    &\times\Bigl[1-R^2e^{2i\delta_1\tau}-R^2e^{2i(\delta_1+\Omega)\tau}+
    R^2e^{2i(2\delta_1+\Omega)\tau}\Bigr],
\end{align*}
describes the difference between $a^{\textrm{r}}_{\textrm{out}}$ and
$a^{\textrm{t}}_{\textrm{out}}$, playing the key role in our further
consideration. In the resonant regime ($\delta_1=0$) we have
$\sigma_1=1$, thus it is convenient to rewrite factor $\sigma_1$ as
a sum $1+\Delta\sigma_1$, where:
\begin{equation*}
    \Delta\sigma_1=\bigl(1-e^{2i\delta_1\tau}\bigr)\,
    \frac{1-R^2e^{2i(\delta_1+\Omega)\tau}}{T^2}\,e^{-2i\delta_1\tau},
\end{equation*}
Remind also, that the transmitted wave is redirected towards
platform $\textrm{P}_1$ for detection. Therefore, the truly measured
quantity is
$a^{\textrm{t}}_{\textrm{out}}e^{i(\delta_1+\Omega)\tau}$. However,
keeping this in mind, below we deal only with
$a^{\textrm{t}}_{\textrm{out}}$. The additional phase can be taken
into account straightforwardly.

We should now express the obtained result in terms of (i) the
fluctuative displacements measured in the laboratory frame and (ii)
the GW displacement measured in the local Lorentz frame of platform
$\textrm{P}_1$. According to formula (\ref{law_of_motion}) the
transformation law is:
\begin{subequations}
\begin{align}
    X_a(t)&=\xi_a(t)-\xi_{\textrm{P}_1}(t),\label{X_a}\\
    X_b(t)&=\frac{1}{2}\,Lh(t)+\xi_b(t)-\xi_{\textrm{P}_1}(t).\label{X_b}
\end{align}
\end{subequations}
Here we denoted the fluctuative motions of mirrors $a$ and $b$ as
$\xi_{a,b}$. These formulas are strict for any separation between
the mirrors. Substituting $X_a$ and $X_b$ into the response signals
(\ref{reflected_signal}) and (\ref{transmitted_signal}) we rewrite
them in terms of the GW signal
\begin{equation*}
    \xi_{\textrm{gw}}(\Omega)=\frac{1}{2}\,Lh(\Omega)\,\frac{\sin\Omega\tau}{\Omega\tau},
\end{equation*}
and fluctuating displacements $\xi_{a,b,\textrm{P}_1}$:
\begin{subequations}
\begin{align}
    a^{\textrm{r}}_{\textrm{out}}&=
    \mathcal{R}_1a_{\textrm{in}}+\mathcal{T}_1a_{\textrm{vac}}\nonumber\\
    &-\frac{RT^2A_{\textrm{in}0}e^{2i\delta_1\tau}}
    {\mathcal{T}^2_{\delta_1}\mathcal{T}^2_{\delta_1+\Omega}}\,
    2ik_1\Bigl[\xi_be^{i\Omega\tau}-\sigma_1\xi_a+
    \xi_{\textrm{gw}}e^{i\Omega\tau}\Bigr]\nonumber\\
    &-\frac{RT^2A_{\textrm{in}0}e^{2i\delta_1\tau}}
    {\mathcal{T}^2_{\delta_1}\mathcal{T}^2_{\delta_1+\Omega}}\,
    ik_1\xi_{\textrm{P}_1}\bigl(2\sigma_1-1-e^{2i\Omega\tau}\bigr),
    \label{reflected_signal_gw_fl}\\
    a^{\textrm{t}}_{\textrm{out}}&=
    \mathcal{T}_1a_{\textrm{in}}+\mathcal{R}_1a_{\textrm{vac}}\nonumber\\
    &+\frac{R^2T^2A_{\textrm{in}0}e^{3i\delta_1\tau}}
    {\mathcal{T}^2_{\delta_1}\mathcal{T}^2_{\delta_1+\Omega}}\,
    2ik_1\Bigl[\xi_be^{i\Omega\tau}-\xi_a+
    \xi_{\textrm{gw}}e^{i\Omega\tau}\Bigr]e^{i\Omega\tau}\nonumber\\
    &+\frac{R^2T^2A_{\textrm{in}0}e^{3i\delta_1\tau}}
    {\mathcal{T}^2_{\delta_1}\mathcal{T}^2_{\delta_1+\Omega}}\,
    ik_1\xi_{\textrm{P}_1}\bigl(1-e^{2i\Omega\tau}\bigr)e^{i\Omega\tau}.
    \label{transmitted_signal_gw_fl}
\end{align}
\end{subequations}
The following notations have been introduced above:
\begin{align*}
    \mathcal{T}^2_{\delta_1}&=1-R^2e^{2i\delta_1\tau},\qquad
    \mathcal{T}^2_{\delta_1+\Omega}=1-R^2e^{2i(\delta_1+\Omega)\tau},\\
    \mathcal{R}_1&=
    \frac{R-Re^{2i(\delta_1+\Omega)\tau}}{1-R^2e^{2i(\delta_1+\Omega)\tau}},\qquad
    \mathcal{T}_1=\frac{T^2e^{i(\delta_1+\Omega)\tau}}{1-R^2e^{2i(\delta_1+\Omega)\tau}},
\end{align*}
having the following physical meaning: $1/\mathcal{T}^2_{\delta_1}$
describes the resonant amplification of the input amplitude
$A_{\textrm{in}0}$ inside the cavity,
$1/\mathcal{T}^2_{\delta_1+\Omega}$ describes the
frequency-dependent resonant amplification of the variation of the
circulating light wave, $\mathcal{R}_1$ and $\mathcal{T}_1$ are the
generalized coefficients of reflection (from a FP cavity) and
transmission (through a FP cavity).

It is convenient to analyze the physical meaning of the obtained
formulas. Fist we consider the reflected wave rewriting it in the
following form:
\begin{align*}
    a^{\textrm{r}}_{\textrm{out}}&=
    \mathcal{R}_1(a_{\textrm{in}}-A_{\textrm{in}0}ik_1\xi_{\textrm{P}_1})+
    \mathcal{T}_1a_{\textrm{vac}}\nonumber\\
    &\quad+TA_{-0}2ik_1\Bigl[(\xi_b+\xi_{\textrm{gw}})e^{i\Omega\tau}-\xi_a\Bigr]
    /\mathcal{T}^2_{\delta_1+\Omega}\nonumber\\
    &\quad+A^{\textrm{r}}_{\textrm{out}0}2ik_1\xi_a-
    A^{\textrm{r}}_{\textrm{out}0}ik_1\xi_{\textrm{P}_1}.
\end{align*}
The 1st term states that the optical laser noise $a_{\textrm{in}}$
is indistinguishable from laser displacement noise
$\xi_{\textrm{P}_1}$, so they always come together. The 2nd summand
describes the propagation of the vacuum noise through a FP cavity.
The 3rd term is the light wave flowing out of the cavity containing
the accumulated phase shift. The 4th summand, which is responsible
for $\Delta\sigma_1$, describes the prompt reflection from the input
mirror $a$. The last term describes the phase shift acquired by the
light wave due to displacement noise of detector on platform
$\textrm{P}_1$.

In a similar way one can consider the transmitted wave. The only
difference which should be taken into account is the following: the
term proportional to $\xi_{\textrm{P}_1}$ in formula
(\ref{transmitted_signal_gw_fl}) cannot be reduced to
$-\mathcal{T}_1A_{\textrm{in}0}ik_1\xi_{\textrm{P}_1}$ due to the
detection scheme we use for the transmitted wave. If one adds the
$A_{\textrm{out}0}^{\textrm{t}}$-summand in formula
(\ref{transmitted_wave}) to $a^{\textrm{t}}_{\textrm{out}}$ then
$-\mathcal{T}_1A_{\textrm{in}0}ik_1\xi_{\textrm{P}_1}$ is recovered.

\section{Double-pumped Fabry-Perot cavity}\label{sec_DPFP}
\subsection{Response signals of a double-pumped Fabry-Perot cavity}
\begin{figure}[h]
\begin{center}
\includegraphics[scale=0.58]{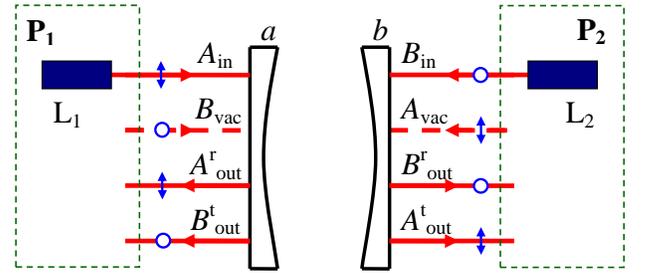}
\caption{Fabry-Perot cavity pumped through both of its mirrors (a
DPFP cavity). Lasers $\textrm{L}_1$ and $\textrm{L}_2$ are rigidly
mounted on platforms $\textrm{P}_1$ and $\textrm{P}_2$ respectively.
The pump wave through mirror $a$ is denoted with $A_{\textrm{in}}$
and is assumed to be polarized in the plane of incidence. The pump
wave through mirror $b$ is denoted with $B_{\textrm{in}}$ and is
assumed to be polarized normally to the plane of incidence.
Corresponding vacuum pumps are $A_{\textrm{vac}}$ and
$B_{\textrm{vac}}$. Output ports are
$A^{\textrm{r,t}}_{\textrm{out}}$ and
$B^{\textrm{r,t}}_{\textrm{out}}$.}\label{pic_DPFP_cavity}
\end{center}
\end{figure}

Let a single Fabry-Perot cavity be pumped through both of its
mirrors (see Fig. \ref{pic_DPFP_cavity}). We assume the pump wave
through mirror $a$ to have amplitude $\mathcal{A}$, detuning
$\delta_1$ (carrier frequency $\omega_1$), polarization in the plane
of incidence and denote it with $A_{\textrm{in}}$; the pump wave
through mirror $b$ is assumed to have amplitude $\mathcal{B}$,
detuning $\delta_2$ (carrier frequency $\omega_2$), polarization
orthogonal to the plane of incidence and is denoted with
$B_{\textrm{in}}$. Corresponding vacuum pumps through mirrors $b$
and $a$ are denoted with $A_{\textrm{vac}}$ and $B_{\textrm{vac}}$.

The response functions corresponding to the pump through mirror $b$
are straightforwardly obtained from functions
(\ref{reflected_signal_gw_fl}, \ref{transmitted_signal_gw_fl})
replacing $\delta_1\rightarrow\delta_2$, $\xi_a\rightarrow-\xi_b$,
$\xi_b\rightarrow-\xi_a$,
$\xi_{\textrm{P}_1}\rightarrow-\xi_{\textrm{P}_2}$ and keeping the
GW term unchanged due to the symmetry of the system and plane GW
wavefront. For convenience we gather signals in all the four output
ports of the DPFP cavity omitting spectral arguments and taking into
account the relation $k_1\approx k_2\equiv k_0$ valid for the
corresponding carrier frequencies $\omega_1$ and $\omega_2$ lying
within the same resonance curve:
\begin{subequations}
\begin{align}
    a^{\textrm{r}}_{\textrm{out}}&=\mathcal{R}_1a_{\textrm{in}}+
    \mathcal{T}_1a_{\textrm{vac}}\nonumber\\
    &-\frac{RT^2\mathcal{A}e^{2i\delta_1\tau}}
    {\mathcal{T}^2_{\delta_1}\mathcal{T}^2_{\delta_1+\Omega}}\,
    2ik_0\Bigl[(\xi_b+\xi_{\textrm{gw}})e^{i\Omega\tau}-
    \sigma_1\xi_a\Bigr]\nonumber\\
    &-\frac{RT^2\mathcal{A}e^{2i\delta_1\tau}}
    {\mathcal{T}^2_{\delta_1}\mathcal{T}^2_{\delta_1+\Omega}}\,
    ik_0\xi_{\textrm{P}_1}\bigl(2\sigma_1-1-e^{2i\Omega\tau}\bigr),
    \label{a_r}\\
    a^{\textrm{t}}_{\textrm{out}}&=\mathcal{T}_1a_{\textrm{in}}+
    \mathcal{R}_1a_{\textrm{vac}}\nonumber\\
    &+\frac{R^2T^2\mathcal{A}e^{3i\delta_1\tau}}
    {\mathcal{T}^2_{\delta_1}\mathcal{T}^2_{\delta_1+\Omega}}\,
    2ik_0\Bigl[(\xi_b+\xi_{\textrm{gw}})e^{2i\Omega\tau}-
    \xi_ae^{i\Omega\tau}\Bigr]\nonumber\\
    &+\frac{R^2T^2\mathcal{A}e^{3i\delta_1\tau}}
    {\mathcal{T}^2_{\delta_1}\mathcal{T}^2_{\delta_1+\Omega}}\,
    ik_0\xi_{\textrm{P}_1}\bigl(1-e^{2i\Omega\tau}\bigr)e^{i\Omega\tau},
    \label{a_t}\\
    b^{\textrm{r}}_{\textrm{out}}&=\mathcal{R}_2b_{\textrm{in}}+
    \mathcal{T}_2b_{\textrm{vac}}\nonumber\\
    &-\frac{RT^2\mathcal{B}e^{2i\delta_2\tau}}
    {\mathcal{T}^2_{\delta_2}\mathcal{T}^2_{\delta_2+\Omega}}\,
    2ik_0\Bigl[(-\xi_a+\xi_{\textrm{gw}})e^{i\Omega\tau}+
    \sigma_2\xi_b\Bigr]\nonumber\\
    &+\frac{RT^2\mathcal{B}e^{2i\delta_2\tau}}
    {\mathcal{T}^2_{\delta_2}\mathcal{T}^2_{\delta_2+\Omega}}\,
    ik_0\xi_{\textrm{P}_2}\bigl(2\sigma_2-1-e^{2i\Omega\tau}\bigr),
    \label{b_r}\\
    b^{\textrm{t}}_{\textrm{out}}&=\mathcal{T}_2b_{\textrm{in}}+
    \mathcal{R}_2b_{\textrm{vac}}\nonumber\\
    &+\frac{R^2T^2\mathcal{B}e^{3i\delta_2\tau}}
    {\mathcal{T}^2_{\delta_2}\mathcal{T}^2_{\delta_2+\Omega}}\,
    2ik_0\Bigl[(-\xi_a+\xi_{\textrm{gw}})e^{2i\Omega\tau}+
    \xi_be^{i\Omega\tau}\Bigr]\nonumber\\
    &-\frac{R^2T^2\mathcal{B}e^{3i\delta_2\tau}}
    {\mathcal{T}^2_{\delta_2}\mathcal{T}^2_{\delta_2+\Omega}}\,
    ik_0\xi_{\textrm{P}_2}\bigl(1-e^{2i\Omega\tau}\bigr)e^{i\Omega\tau}.
    \label{b_t}
\end{align}
\end{subequations}
Here quantities $\mathcal{R}$, $\mathcal{T}$,
$\mathcal{T}^2_{\delta}$ and $\mathcal{T}^2_{\delta+\Omega}$ with
the subscripts ``1'' and ``2'' are evaluated for detunings
$\delta_1$ and $\delta_2$ correspondingly.

The quadrature components of field amplitudes (\ref{b_r}) and
(\ref{b_t}) can be measured in a way similar to the case of a
single-pumped FP cavity (corresponding to field amplitudes
(\ref{a_r}) and (\ref{a_t})). The detection scheme of a DPFP cavity
will require two more homodyne detectors to measure the output
signals corresponding to the second pump.

\subsection{Cancelation of displacement noise}\label{sec_algorithm}
Now we will demonstrate the noise cancelation from the combination
of field amplitudes (\ref{a_r} -- \ref{b_t}). Though it is obvious
and enough from the theoretical point of view, such a consideration
is surely insufficient for the experimental purposes, because we can
only measure quadrature components of the fields, not the complex
field amplitudes themselves. However, we will not present the bulky
calculations of the quadratures here since we consider only
theoretical model, not the specific experimental design.

Therefore, let us assume that we are able to produce any desired
linear combination of the response signals (\ref{a_r} -- \ref{b_t}).
Physically this means that we are able to construct a set of optical
lossless filters with the predetermined transmittance coefficients,
transmit each wave through its filter and then make the waves
interfere.

To illustrate our method of noise elimination we will construct the
linear combination of responses which cancels fluctuating
displacements $\xi_{a,b}$ in three steps. Remind that the
transmitted signals do not take into account the
$e^{i(\delta_{1,2}+\Omega)\tau}$ multiplier.

From the first pair of signals $a^{\textrm{r,t}}_{\textrm{out}}$ we
can eliminate either $\xi_a$ or $\xi_b+\xi_{\textrm{gw}}$. Let us
cancel $\xi_a$. Multiplying $a^{\textrm{r}}_{\textrm{out}}$ on
$Re^{i(\delta_1+\Omega)\tau}$ and adding it to
$\sigma_1a^{\textrm{t}}_{\textrm{out}}$ we obtain:
\begin{align}
    s_1&=Re^{i(\delta_1+\Omega)\tau}a^{\textrm{r}}_{\textrm{out}}+
    \sigma_1a^{\textrm{t}}_{\textrm{out}}\nonumber\\
    &=s_1^{\textrm{fl}}+\frac{R^2T^2\mathcal{A}e^{3i\delta_1\tau}}
    {\mathcal{T}^2_{\delta_1}\mathcal{T}^2_{\delta_1+\Omega}}\,
    2ik_0\Delta\sigma_1(\xi_b+\xi_{\textrm{gw}})e^{2i\Omega\tau}\nonumber\\
    &\qquad-\frac{R^2T^2\mathcal{A}e^{3i\delta_1\tau}}
    {\mathcal{T}^2_{\delta_1}\mathcal{T}^2_{\delta_1+\Omega}}\,
    ik_0\Delta\sigma_1\xi_{P_1}\bigl(1+e^{2i\Omega\tau}\bigr)e^{i\Omega\tau}\nonumber\\
    &=s_1^{\textrm{fl}}+R^2e^{i\delta_1\tau}\bigl(1-e^{2i\delta_1\tau}\bigr)
    \,\frac{\mathcal{A}}{\mathcal{T}^2_{\delta_1}}
    \,2ik_0(\xi_b+\xi_{\textrm{gw}})e^{2i\Omega\tau}\nonumber\\
    &\qquad-R^2e^{i\delta_1\tau}\bigl(1-e^{2i\delta_1\tau}\bigr)
    \,\frac{\mathcal{A}}{\mathcal{T}^2_{\delta_1}}\,
    ik_0\xi_{\textrm{P}_1}\bigl(1+e^{2i\Omega\tau}\bigr)e^{i\Omega\tau},
    \label{no_a}\\
    s_1^{\textrm{fl}}&=a_{\textrm{in}}e^{-i(\delta_1-\Omega)\tau}\nonumber\\
    &\quad+\frac{R}{T^2}\Bigl[e^{2i\Omega\tau}(e^{2i\delta_1\tau}-1)+
    \mathcal{T}^2_{\delta_1}e^{-2i\delta_1\tau}\Bigr]a_{\textrm{vac}}.\nonumber
\end{align}

Similarly, from the second pair of signals
$b^{\textrm{r,t}}_{\textrm{out}}$ we can eliminate either $\xi_b$ or
$-\xi_a+\xi_{\textrm{gw}}$. Since we have already canceled $\xi_a$
from the first pair and are left only with
$\xi_b+\xi_{\textrm{gw}}$, we need to exclude
$-\xi_a+\xi_{\textrm{gw}}$ from the second pair to be left with
$\xi_b$ only. Multiplying $b^{\textrm{r}}_{\textrm{out}}$ on
$Re^{i(\delta_2+\Omega)\tau}$ and adding it to
$b^{\textrm{t}}_{\textrm{out}}$ we obtain:
\begin{align}
    s_2&=Re^{i(\delta_2+\Omega)\tau}b^{\textrm{r}}_{\textrm{out}}+
    b^{\textrm{t}}_{\textrm{out}}\nonumber\\
    &=s_2^{\textrm{fl}}-\frac{R^2T^2\mathcal{B}e^{3i\delta_2\tau}}
    {\mathcal{T}^2_{\delta_2}\mathcal{T}^2_{\delta_2+\Omega}}\,
    2ik_0\Delta\sigma_2(\xi_b-\xi_{\textrm{P}_2})e^{i\Omega\tau}\nonumber\\
    &=s_2^{\textrm{fl}}-R^2e^{i\delta_2\tau}\bigl(1-e^{2i\delta_2\tau}\bigr)
    \,\frac{\mathcal{B}}{\mathcal{T}^2_{\delta_2}}\,
    2ik_0(\xi_b-\xi_{\textrm{P}_2})e^{i\Omega\tau},
    \label{no_a_gw}\\
    s_2^{\textrm{fl}}&=b_{\textrm{in}}e^{i(\delta_2+\Omega)\tau}+
    Rb_{\textrm{vac}}\nonumber.
\end{align}

To perform the last step we need to introduce the relation between
$\mathcal{A}$ and $\mathcal{B}$. It is convenient (but not
necessary) to assume
$\mathcal{A}/\mathcal{T}^2_{\delta_1}=\mathcal{B}/\mathcal{T}^2_{\delta_2}$.
Ultimately we cancel the information about $\xi_b$ from the pair of
signals $s_{1,2}$:
\begin{align}
    s&=s_1+
    \frac{e^{i\delta_1\tau}\bigl(1-e^{2i\delta_1\tau}\bigr)}
    {e^{i\delta_2\tau}\bigl(1-e^{2i\delta_2\tau}\bigr)}\,
    s_2e^{i\Omega\tau}\nonumber\\
    &=s^{\textrm{fl}}+R^2e^{i(\delta_1+\Omega)\tau}\bigl(1-e^{2i\delta_1\tau}\bigr)
    \frac{\mathcal{A}}{\mathcal{T}^2_{\delta_1}}\nonumber\\
    &\qquad\qquad\times ik_0\Bigl[-\xi_{\textrm{P}_1}+2(\xi_{\textrm{P}_2}+\xi_{\textrm{gw}})
    e^{i\Omega\tau}-\xi_{\textrm{P}_1}e^{2i\Omega\tau}\Bigr],
    \label{DFI_signal}\\
    s^{\textrm{fl}}&=a_{\textrm{in}}e^{-i(\delta_1-\Omega)\tau}+
    \frac{1-e^{2i\delta_1\tau}}{1-e^{2i\delta_2\tau}}\,
    b_{\textrm{in}}e^{i(\delta_1+2\Omega)\tau}\nonumber\\
    &\quad+\frac{R}{T^2}\Bigl[e^{2i\Omega\tau}\bigl(e^{2i\delta_1\tau}-1\bigr)+
    \mathcal{T}^2_{\delta_1}e^{-2i\delta_1\tau}\Bigr]a_{\textrm{vac}}\nonumber\\
    &\quad+\frac{e^{i\delta_1\tau}\bigl(1-e^{2i\delta_1\tau}\bigr)}
    {e^{i\delta_2\tau}\bigl(1-e^{2i\delta_2\tau}\bigr)}\,
    Rb_{\textrm{vac}}e^{i\Omega\tau}.\nonumber
\end{align}
Total signal $s$, below called DFI response signal, does not contain
information about displacement noise of the mirrors but is not free
from displacement noise of the platforms.

For the ground-based detectors with the spacial scale $L$ of several
kilometers the most important is the low-frequency response, i.e.
the limit $\Omega L/c\ll1$. We will analyze two special cases.

In the simplest case of equal pumps we have
$\mathcal{A}=\mathcal{B}$ and $\delta_1=\delta_2$. Then in the
narrow-band approximation ($T^2=2\gamma\tau\ll1$,
$\delta_{1,2}\tau\ll1$, where $\gamma$ is the cavity
half-bandwidth):
\begin{align}
    s|_{\delta_2=\delta_1}&\approx
    a_{\textrm{in}}+b_{\textrm{in}}+a_{\textrm{vac}}+b_{\textrm{vac}}\nonumber\\
    &\quad-\frac{i\delta_1}{\gamma-i\delta_1}\,
    \mathcal{A}\,2ik_0\left(\frac{1}{2}\,Lh+\xi_{\textrm{P}_2}-\xi_{\textrm{P}_1}\right).
    \label{DFI_equal_pumps_approx}
\end{align}

Remind \cite{1967_EM_rigidity,1997_optical_bars,1999_rigidity_QM,
2000_rakh_phd_thesis,2001_FD_rigidity}, that due to the significant
amplification of the input laser power inside a FP cavity test
masses are subjected to the force of radiation pressure. It is known
that the sign of the induced ponderomotive rigidity depends on the
sign of detuning. Therefore, in order to cancel the effects of
radiation pressure we should consider the pumps with opposite
detunings $\delta_2=-\delta_1$. In this case both the pumps create
ponderomotive rigidities with the opposite signs and total rigidity
vanishes. The DFI signal in this case is:
\begin{align}
    s|_{\delta_2=-\delta_1}&\approx
    a_{\textrm{in}}-b_{\textrm{in}}+a_{\textrm{vac}}-b_{\textrm{vac}}\nonumber\\
    &\quad-\frac{i\delta_1}{\gamma-i\delta_1}\,
    \mathcal{A}\,2ik_0\left(\frac{1}{2}\,Lh+\xi_{\textrm{P}_2}-\xi_{\textrm{P}_1}\right).
    \label{DFI_diff_pumps_approx}
\end{align}

Obviously, in the previous case of equal detunings total
ponderomotive rigidity does not vanish and, strictly speaking, the
effects of radiation pressure in the DPFP cavity require separate
detailed analysis.

From formulas (\ref{DFI_equal_pumps_approx}) and
(\ref{DFI_diff_pumps_approx}) we conclude that the signal-to-noise
ratio of the DPFP cavity operating as the displacement-noise-free
detector is of the same order as for the configuration with two test
masses and only one round trip of light between them (i.e. without
the resonant gain).

\section{Discussion}\label{sec_discussion}
Let us now discuss several issues concerning the noise cancelation
in the proposed model.

\subsection{Special cases}
First, it is useful to consider two special cases when noise
cancelation is impossible.
\begin{enumerate}
\item Resonant pump. One can derive from formula
(\ref{reflected_signal_gw_fl}) that the coefficient $p$ in formula
(\ref{simple_refl_sig}) is proportional to the amplitude of
reflected wave $A^{\textrm{r}}_{\textrm{out}0}$. In Appendix
\ref{app_boundaries} it is found that
$A^{\textrm{r}}_{\textrm{out}0}=RA_{\textrm{in}0}(1-e^{2i\delta_1\tau})/\mathcal{T}^2_{\delta_1}$.
Thus in the resonant regime ($\delta_1=0$) reflected wave has no
``strong'' component meaning that the prompt reflection from the
input mirror does not occur and $p=0$. As a result, both the
reflected and transmitted signals become indistinguishable, i.e.
they carry equal amount of information about the coordinates of the
mirrors (see equations (\ref{simple_trans_sig}) and
(\ref{simple_refl_sig})). In general case (formulas
(\ref{reflected_signal_gw_fl}) and (\ref{transmitted_signal_gw_fl}))
the resonant regime corresponds to $\Delta\sigma_1=0$, resulting in
the relation
$a^{\textrm{t}}_{\textrm{out}}=-Ra^{\textrm{r}}_{\textrm{out}}e^{i\Omega\tau}$,
neglecting the optical noise.
\item Mirrors mounted on the platforms. One may think of mounting
the mirrors on the platforms to reduce the additional fluctuative
degrees of freedom associated with the platforms. For instance, if
the mirror $a$ is mounted on platform $\textrm{P}_1$ then
$\xi_a=\xi_{\textrm{P}_1}$ and from equation (\ref{simple_refl_sig})
it is evident that both the responses become equivalent. In general
case (see formulas (\ref{reflected_signal}) and
(\ref{transmitted_signal}) it is evident that for
$X_a=\xi_a-\xi_{\textrm{P}_1}=0$ again
$a^{\textrm{t}}_{\textrm{out}}=-Ra^{\textrm{r}}_{\textrm{out}}e^{i\Omega\tau}$.
\end{enumerate}

\subsection{Optical power requirements}
The loss of the resonant gain in a DPFP cavity also results in
increase of the optical power needed to reach the SQL level of
sensitivity. In conventional (LIGO) topology both the mean amplitude
and the signal are resonantly amplified resulting in less power
needed to reach SQL as compared to any single-round-trip detector.
For instance, in Advanced LIGO detectors (utilizing also the power
recycling mirrors) SQL will be reached with $\approx 1$ MW of
circulating optical power corresponding to $\approx 100$ W laser. In
contrast, in a DPFP cavity the same level of sensitivity will be
reached at $\approx 1$ GW of laser power. This number might not seem
so dramatic if one reminds that the squeezed light allows to
decrease the power needed. To achieve the high factors of squeezing
one must provide the mirrors with the coefficient of optical losses
as small as possible; according to J.M. Makowsky there is a strong
evidence that the loss coefficient $\sim 10^{-9}$ will be reached in
the near future.

\subsection{Limitations due to the relativity principle}
Remind that in formula (\ref{simple_refl_sig})
$\xi_{\textrm{gw}}\approx Lh/2$ (see also formulas
(\ref{DFI_equal_pumps_approx}) and (\ref{DFI_diff_pumps_approx})),
thus direct coupling of the GW to the light wave plays no role in
our noise-cancelation scheme. From the obtained results (see also
reasonings in Sec. \ref{sec_mechanism}) it seems that it is hardly
possible (without contradicting the relativity principle) to
completely eliminate the displacement noise, keeping simultaneously
the $h(L/\lambda_{\textrm{gw}})^0$ or $h(L/\lambda_{\textrm{gw}})^1$
order of the DFI signal, since these orders correspond to coordinate
and velocity measurements. Relativity principle forbids absolute
coordinate and velocity measurements; only acceleration, in
principle, can be measured absolutely, corresponding to complete DFI
of the $h(L/\lambda_{\textrm{gw}})^n$, $n\geq2$ order proposed by
Kawamura \textit{et al.} Thus we are left to choose either sacrifice
with the GW sensitivity but completely eliminate displacement noise,
or keep good GW sensitivity at the expense of incomplete noise
cancelation. To suppress the fluctuations associated with the
platforms (where lasers and detectors are mounted) one will need to
increase their masses and cool them down to cryogenic temperatures.
The only limiting factors will be left then are the classical
(laser) and vacuum optical noises.

\subsection{Further prospects: cancelation of laser noise and detection schemes}
\label{sec_prospects} 
\begin{figure}[h]
\begin{center}
\includegraphics[scale=0.59]{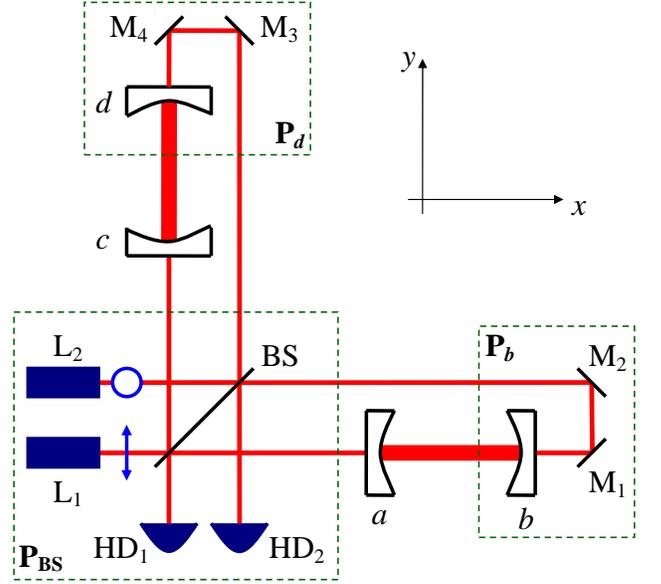}
\caption{A Michelson/DPFP optical setup. DPFP cavities $ab$ and $cd$
are inserted into the horizontal and vertical arms of Michelson
interferometer correspondingly. Lasers $\textrm{L}_1$ and
$\textrm{L}_2$, beamsplitter BS and homodyne detectors
$\textrm{HD}_1$ and $\textrm{HD}_2$ are rigidly mounted on platform
$\textrm{P}_{\textrm{BS}}$. Cavity mirror $b$ and auxiliary mirrors
$\textrm{M}_1$ and $\textrm{M}_2$ are rigidly mounted on platform
$\textrm{P}_b$; cavity mirror $d$ and auxiliary mirrors
$\textrm{M}_3$ and $\textrm{M}_4$ are rigidly mounted on platform
$\textrm{P}_d$. Detector $\textrm{HD}_1$ measures the quadratures of
reflected wave corresponding to laser $\textrm{L}_1$ and the ones of
transmitted wave corresponding to laser $\textrm{L}_2$. Detector
$\textrm{HD}_2$ measures the quadratures of reflected wave
corresponding to laser $\textrm{L}_2$ and the ones of transmitted
wave corresponding to laser $\textrm{L}_1$.}
\label{pic_Michelson_DPFP}
\end{center}
\end{figure}

From formula (\ref{DFI_equal_pumps_approx}) or
(\ref{DFI_diff_pumps_approx}) one may conclude that the
\textit{fundamental} limitations of the proposed scheme are (i) the
vacuum shot noise ($a_{\textrm{vac}}$ and $b_{\textrm{vac}}$ terms)
due to the uncertainty principle and (ii) the residual displacement
noise ($\xi_{\textrm{P}_1}$ and $\xi_{\textrm{P}_2}$ terms) due to
the relativity principle as discussed above. It is also known that
laser noise can be eliminated in differential (balanced) optical
setup, for instance Mach-Zehnder or Michelson interferometer. Since
laser noise dominates over vacuum shot noise in practice, one needs
to implement the proposed DPFP cavity into some balanced scheme to
increase the overall SNR. Here we propose one of the obvious
modifications of LIGO topology, namely a Michelson interferometer
with two DPFP cavities in its arms, which utilizes a ``round-trip
ideology'' widely used in this paper (see Fig.
\ref{pic_Michelson_DPFP}).

First, we describe the operation of the scheme as a whole and then
consider noise cancelation issue. Let laser $\textrm{L}_1$ emit the
optical wave polarized in the plane of incidence. Upon arrival to
beamsplitter BS optical wave is splitted into two beams: the one
traveling in the horizontal arm towards FP cavity assembled of
mirrors $a$ and $b$ and the other traveling in the vertical arm
towards FP cavity $cd$. Both reflected waves then reunite at
beamsplitter and the resulting optical field is detected by homodyne
detector $\textrm{HD}_1$. The wave transmitted through $ab$ cavity
is redirected towards beamsplitter by auxiliary mirrors
$\textrm{M}_1$ and $\textrm{M}_2$. Similarly, the wave transmitted
through $cd$ cavity is redirected towards beamsplitter by mirrors
$\textrm{M}_3$ and $\textrm{M}_4$. Ultimately, both transmitted
waves interfere at beamsplitter and are detected by homodyne
detector $\textrm{HD}_2$.

Let the second pump be produced by laser $\textrm{L}_2$ emitting the
radiation polarized normally to the plane of incidence. Input wave
inside the horizontal arm produces the reflected wave via
$\textrm{BS} - \textrm{M}_2 - \textrm{M}_1 - ab\textrm{ cavity} -
\textrm{M}_1 - \textrm{M}_2 - \textrm{BS}$ optical path and the
transmitted wave via $\textrm{BS} - \textrm{M}_2 -  \textrm{M}_1 -
ab\textrm{ cavity} - \textrm{BS}$ path. Similarly, reflected and
transmitted waves are produced in the vertical arm. Interfering
reflected waves are detected then by homodyne detector
$\textrm{HD}_2$ and transmitted waves are detected by
$\textrm{HD}_1$ detector.

Following the consideration of a single DPFP cavity, we may assume
that several optical elements are rigidly attached to each other.
For instance, let us assume that both lasers, beamsplitter and both
detectors are rigidly mounted on platform
$\textrm{P}_{\textrm{BS}}$; mirror $b$ and small auxiliary mirrors
$\textrm{M}_1$ and $\textrm{M}_2$ are mounted on platform
$\textrm{P}_b$; mirror $d$ and mirrors $\textrm{M}_3$ and
$\textrm{M}_4$ are mounted on platform $\textrm{P}_d$. Then there
are left only six essential degrees of freedom: displacement of
$\textrm{P}_{\textrm{BS}}$ along $x$- and $y$-axes, displacements of
$a$ and $\textrm{P}_b$ along the $x$-axis, and displacements of $c$
and $\textrm{P}_d$ along the $y$-axis. Let us denote the coordinate
fluctuations of the $j$th test mass corresponding to the motion
along the $x$- and $y$-axes as $\xi_j$ and $\eta_j$. Each of four
interferometer responses $a_i$ contains displacement noise in the
combinations of the following type:
\begin{multline*}
    a_i\sim a_{\textrm{HD}_1}+a_{\textrm{HD}_2}+
    k_0(\xi_{\textrm{P}_{\textrm{BS}}}-\eta_{\textrm{P}_{\textrm{BS}}})\\
    +k_0\left(\xi_{\textrm{P}_b}-\xi_a+\frac{1}{2}\,Lh\right)
    -k_0\left(\eta_{\textrm{P}_d}-\eta_c-\frac{1}{2}\,Lh\right),
\end{multline*}
where $a_{\textrm{HD}_i}$ is the vacuum shot noise in the dark port
of detector $\textrm{HD}_i$. Note that the terms describing optical
laser noise are absent since it vanishes due to the interference of
the waves at beamsplitter. In fact it is not necessary to demand
that lasers are rigidly attached to beamsplitter: since optical and
displacement noise of a laser are indistinguishable (there sum is
usually called laser phase noise) both noises are canceled
simultaneously. Here we do not calculate explicitly the coefficients
before each noise term in $a_i$ since they depend on specific
details of the optical setup.

Excluding $\xi_{\textrm{P}_b}-\eta_{\textrm{P}_d}$ and
$\xi_a-\eta_c$ from the linear combination of responses one obtains
signal with only fundamental noises (for $(\Omega\tau)^0$-DFI) left:
\begin{equation*}
    s_{\textrm{DFI}}\sim a_{\textrm{HD}_1}+a_{\textrm{HD}_2}+
    k_0(\xi_{\textrm{P}_{\textrm{BS}}}-\eta_{\textrm{P}_{\textrm{BS}}}+Lh).
\end{equation*}

Obviously, the major drawback of the proposed scheme is the
significant amount of additional optical elements such as
beamsplitter and mirrors used to split and redirect laser beams. Our
assumption that several elements could be rigidly installed on the
platforms (i.e. to be noiseless) should be validated in practice.

The related problem is the construction of the most practical
measurement schemes. In particular, when analyzing the transmitted
wave in a single DPFP cavity above, we dealt only with the
round-trip measurement schemes, i.e. redirected the transmitted
radiation for detection towards the location (approximately) of
emitting device. To clarify our analysis we also made the
corresponding reference oscillation to perform a round trip. This
may seem inconvenient (but certainly not impossible) to the
experimentalists, thus other possibilities could be explored. For
instance, one may think of forward-trip measurement schemes
\cite{accel_observ}, i.e. the situation when transmitted wave is
detected straightforwardly (without any redirection). Corresponding
balanced schemes could be proposed then.

\section{Conclusion}
In this paper we have analyzed the operation of a Fabry-Perot cavity
pumped through both the mirrors (a DPFP cavity) performing the
mirrors-displacement-noise-free gravitational-wave detection. We
have demonstrated that due to the asymmetry between the reflection
and transmission output ports of detuned cavity it is possible to
construct a linear combination of four response signals which
cancels displacement fluctuations of the mirrors. At low frequencies
the GW response of the DPFP cavity turns out to be far better than
that of the Mach-Zehnder-based DFIs proposed by S. Kawamura
\textit{et al.} due to the different mechanisms of
noise-cancelation. However, the effective loss of the resonant gain
results in the sensitivity limitation of the DPFP cavity by
displacement noise of lasers and detectors.

The performed analysis suggests that though addressed as a toy model
in this paper, DPFP cavity can be considered a promising candidate
for constituent part of the future generation GW detectors, provided
the noises of lasers and detectors are suppressed: it allows the
significant extension of the frequency band of the ground-based
detectors and by elimination of back-action noise straightforwardly
avoids the standard quantum limitation.

The problems of (i) DPFP-based laser-noise-cancelation schemes, (ii)
practical measurement schemes and (iii) radiation pressure effects
in a DPFP cavity require future investigation. We hope that
presented analysis will stimulate the search for new configurations
of FP-based displacement-noise-free GW detectors.

\acknowledgements We would like to thank V.B. Braginsky, M.L.
Gorodetsky and F.Ya. Khalili for fruitful discussions and valuable
critical remarks on the paper. We are very grateful to T. Corbitt
for his suggestions about the improvement of the text. In
particular, we would like to express our gratitude to Y. Chen for
the hospitality and support during our stay at AEI and the inspiring
discussions which greatly helped to improve our research. This work
was supported by LIGO team from Caltech and in part by NSF and
Caltech grant PHY-0353775 and by Grant of President of Russian
Federation NS-5178.2006.2.

\appendix
\section{Quantized electromagnetic wave}\label{app_quant_emw}
In this Appendix we introduce the notations for the quantized field
of electromagnetic wave which will be used throughout the paper.

In quantum electrodynamics the operator of electric field in
Heisenberg picture is:
\begin{equation*}
    A(x,t)=\int_0^\infty\sqrt{\frac{2\pi\hbar\omega}{Sc}}\,a(\omega)
    e^{-i\omega(t-x/c)}\,\frac{d\omega}{2\pi}+{\textrm{h.c.}},
\end{equation*}
where $S$ is the effective cross section area of the laser beam and
$a(\omega)$ is the annihilation operator obeying the commutation
relations
\begin{equation*}
    \bigl[a(\omega),a(\omega')\bigr]=0,\quad
    \bigl[a(\omega),a^\dag(\omega')\bigr]=2\pi\delta(\omega-\omega').
\end{equation*}

It will be convenient now to introduce the carrier frequency
$\omega_0$: $\omega=\omega_0+\Omega$, $|\Omega|\ll\omega_0$, and to
rewrite the field operator in the following way:
\begin{multline*}
    A(x,t)=e^{-i(\omega_0t-k_0x)}\\
    \times\int_{-\omega_0}^\infty
    \sqrt{\frac{2\pi\hbar(\omega_0+\Omega)}{Sc}}\,a(\omega_0+\Omega)
    e^{-i\Omega(t-x/c)}\,\frac{d\Omega}{2\pi}+{\textrm{h.c.}},
\end{multline*}
where $k_0=\omega_0/c$. Now we split the annihilation operator into
two summands:
\begin{equation*}
    a(\omega_0+\Omega)=A_0\delta(0)+a'(\omega_0+\Omega).
\end{equation*}
For convenience we change notation $a'\rightarrow a$ since we do not
need old $a$ any further. Extending now the lower limit of
integration to $-\infty$ (since $|\Omega|\ll\omega_0$), we finally
obtain the double-sided (from $-\infty$ to $+\infty$) expression for
the field operator:
\begin{multline}
    A(x,t)=\sqrt{\frac{2\pi\hbar\omega_0}{Sc}}\,e^{-i(\omega_0t-k_0x)}\\
    \times\left[A_0+\int_{-\infty}^{+\infty}
    a(\omega_0+\Omega)
    e^{-i\Omega(t-x/c)}\,\frac{d\Omega}{2\pi}\right]
    +{\textrm{h.c.}}\label{electric field}
\end{multline}
In these notations electric field of the wave is represented as a
sum of (i) ``strong'' (classical) wave with amplitude $A_0$ and
(carrier) frequency $\omega_0$ and (ii) ``weak'' wave describing the
quantum fluctuations of the optical field with its amplitude obeying
the commutation relations:
\begin{align*}
    &\bigl[a(\omega_0+\Omega),a(\omega_0+\Omega')\bigr]=0,\\
    &\bigl[a(\omega_0+\Omega),a^\dag(\omega_0+\Omega')\bigr]=2\pi\delta(\Omega-\Omega').
\end{align*}
The double-sided expression is the one most close to the Fourier
representation of the classical fields and will be used throughout
the paper. For convenience we omit the
$\sqrt{2\pi\hbar\omega_0/Sc}$-multiplier in the main body of the
paper since it is the common multiplier in all the equations.

For completeness we also introduce the quadrature components of the
wave. Formula (\ref{electric field}) can be rewritten as:
\begin{align}
    &A(x,t)=\sqrt{\frac{2\pi\hbar\omega_0}{Sc}}\,e^{-i(\omega_0t-k_0x)}\nonumber\\
    &\times\left\{A_0+\int_0^\infty
    \Bigl[a_{\omega_0+\Omega}e^{-i\Omega(t-x/c)}+
    a_{\omega_0-\Omega}e^{i\Omega(t-x/c)}\Bigr]\frac{d\Omega}{2\pi}\right\}\nonumber\\
    &+{\textrm{h.c.}},
    \label{electric field_single_photon}
\end{align}
where $a_{\omega_0-\Omega}$ obeys the same commutation relation as
$a_{\omega_0+\Omega}$:
\begin{equation*}
    [a_{\omega_0+\Omega},a_{\omega_0+\Omega'}^\dag]=
    [a_{\omega_0-\Omega},a_{\omega_0-\Omega'}^\dag]=2\pi\delta(\Omega-\Omega').
\end{equation*}

Next we introduce the so-called correlated two-photon modes with
field operators \cite{1985_two_photon_1,1985_two_photon_2}
\begin{equation*}
    a_\textrm{c}(\Omega)=\frac{a_{\omega_0+\Omega}+a_{\omega_0-\Omega}^\dag}{\sqrt{2}},\qquad
    a_\textrm{s}(\Omega)=\frac{a_{\omega_0+\Omega}-a_{\omega_0-\Omega}^\dag}{\sqrt{2}i},
\end{equation*}
with the only non-zero commutators
\begin{equation*}
    [a_\textrm{c},a_{\textrm{s}'}^\dag]=[a_{\textrm{c}'},a_{\textrm{s}}^\dag]=2\pi
    i\delta(\Omega-\Omega'),
\end{equation*}
where prime denotes the argument with $\Omega'$. In terms of these
two-photon modes formula (\ref{electric field_single_photon}) takes
the form:
\begin{multline*}
    A(x,t)=\sqrt{\frac{4\pi\hbar\omega_0}{Sc}}\,
    \biggl[\sqrt{2}A_0\cos(\omega_0t-k_0x)\\
    +a_{\textrm{c}}(x,t)\cos(\omega_0t-k_0x)
    +a_{\textrm{s}}(x,t)\sin(\omega_0t-k_0x)\biggr],
\end{multline*}
where operators
\begin{align*}
    a_{\textrm{c}}(x,t)&=\int_0^\infty
    a_{\textrm{c}}(\Omega)e^{-i\Omega(t-x/c)}\,\frac{d\Omega}{2\pi}
    +{\textrm{h.c.}},\\
    a_{\textrm{s}}(x,t)&=\int_0^\infty
    a_{\textrm{s}}(\Omega)e^{-i\Omega(t-x/c)}\,\frac{d\Omega}{2\pi}
    +{\textrm{h.c.}},
\end{align*}
in the case $A_0=0$ are called the cosine and sine quadratures (or
quadrature components) correspondingly.


\section{Boundary conditions}\label{app_boundaries}
In this Appendix we solve the set of equations (\ref{bound_a_1} --
\ref{bound_b_2}).

First we substitute fields (\ref{input_wave} --
\ref{transmitted_wave}) into this set and separate the 0th and the
1st order sets.

The zeroth order set is:
\begin{align*}
    A_{+0}&=TA_{\textrm{in}0}-RA_{-0},\\
    A^{\textrm{r}}_{\textrm{out}0}&=RA_{\textrm{in}0}+TA_{-0},\\
    A_{-0}&=-RA_{+0}e^{2i\omega_1\tau},\\
    A^{\textrm{t}}_{\textrm{out}0}&=TA_{+0}e^{i\omega_1\tau}.
\end{align*}
Corresponding solution is:
\begin{align*}
    A_{+0}&=\frac{T}{1-R^2e^{2i\omega_1\tau}}\,A_{\textrm{in}0},\\
    A_{-0}&=-\frac{RTe^{2i\omega_1\tau}}{1-R^2e^{2i\omega_1\tau}}\,A_{\textrm{in}0},\\
    A^{\textrm{t}}_{\textrm{out}0}
    &=\frac{T^2e^{i\omega_1\tau}}{1-R^2e^{2i\omega_1\tau}}\,A_{\textrm{in}0},\\
    A^{\textrm{r}}_{\textrm{out}0}
    &=\frac{R-Re^{2i\omega_1\tau}}{1-R^2e^{2i\omega_1\tau}}\,A_{\textrm{in}0}.
\end{align*}
Amplitudes $A_{\textrm{in}0}$, $A_{\pm0}$ and
$A^{\textrm{r}}_{\textrm{out}0}$ are evaluated at point $x=0$ and
amplitude $A^{\textrm{t}}_{\textrm{out}}$ at point $x=L$.

The first order solution in spectral domain is:
\begin{align*}
    a_+&=Ta_{\textrm{in}}-Ra_-+RA_{-0}2ik_1X_a,\\
    a^{\textrm{r}}_{\textrm{out}}&=Ra_{\textrm{in}}+Ta_-+RA_{\textrm{in}0}2ik_1X_a,\\
    a_-&=Ta_{\textrm{vac}}e^{i(\omega_1+\Omega)\tau}-Ra_+e^{2i(\omega_1+\Omega)\tau}\\
    &\quad-RA_{+0}e^{2i\omega_1\tau}
    \Bigl[2ik_1X_b+g_+(L)-g_-(L)\\
    &\qquad\qquad\qquad\qquad+w_+(L)-w_-(L)\Bigr]e^{i\Omega\tau},\\
    a^{\textrm{t}}_{\textrm{out}}&=Ra_{\textrm{vac}}+Ta_+e^{i(\omega_1+\Omega)\tau}.
\end{align*}
Here $a_i=a_i(\omega_1+\Omega)$, $g_\pm(x)=g_\pm(x,\omega_1+\Omega)$
and $X_i=X_i(\Omega)$. Spectral amplitudes $a_{\textrm{in}}$,
$a_{\pm}$ and $a^{\textrm{r}}_{\textrm{out}}$ are evaluated at point
$x=0$ and amplitude $a^{\textrm{t}}_{\textrm{out}}$ at point $x=L$.
The first order solution is:
\begin{widetext}
\begin{subequations}
\begin{align*}
    a_+&=
    \frac{T}{1-R^2e^{2i(\omega_1+\Omega)\tau}}\,a_{\textrm{in}}-
    \frac{RTe^{i(\omega_1+\Omega)\tau}}{1-R^2e^{2i(\omega_1+\Omega)\tau}}\,a_{\textrm{vac}}+
    \frac{R^2A_{+0}e^{2i\omega_1\tau}}{1-R^2e^{2i(\omega_1+\Omega)\tau}}
    \,i\biggl[2k_1\Bigl(X_be^{i\Omega\tau}-X_a\Bigr)+
    \delta\Psi_{\textrm{emw}}\biggr],\\
    a_-&=
    -\frac{RTe^{2i(\omega_1+\Omega)\tau}}{1-R^2e^{2i(\omega_1+\Omega)\tau}}
    \,a_{\textrm{in}}+
    \frac{Te^{i(\omega_1+\Omega)\tau}}{1-R^2e^{2i(\omega_1+\Omega)\tau}}\,a_{\textrm{vac}}
    +\frac{A_{-0}}{1-R^2e^{2i(\omega_1+\Omega)\tau}}\,i
    \biggl[2k_1\Bigl(X_be^{i\Omega\tau}-\rho_1X_a\Bigr)+
    \delta\Psi_{\textrm{emw}}\biggr],\\
    a^{\textrm{t}}_{\textrm{out}}&=
    \frac{T^2e^{i(\omega_1+\Omega)\tau}}{1-R^2e^{2i(\omega_1+\Omega)\tau}}\,
    a_{\textrm{in}}
    +\frac{R-Re^{2i(\omega_1+\Omega)\tau}}{1-R^2e^{2i(\omega_1+\Omega)\tau}}\,
    a_{\textrm{vac}}
    +\frac{R^2A^{\textrm{t}}_{\textrm{out}0}e^{2i\omega_1\tau}}
    {1-R^2e^{2i(\omega_1+\Omega)\tau}}\,i\biggl[2k_1\Bigl(X_be^{i\Omega\tau}-X_a\Bigr)+
    \delta\Psi_{\textrm{emw}}\biggr]e^{i\Omega\tau},\\
    a^{\textrm{r}}_{\textrm{out}}&=
    \frac{R-Re^{2i(\omega_1+\Omega)\tau}}{1-R^2e^{2i(\omega_1+\Omega)\tau}}\,
    a_{\textrm{in}}
    +\frac{T^2e^{i(\omega_1+\Omega)\tau}}{1-R^2e^{2i(\omega_1+\Omega)\tau}}\,
    a_{\textrm{vac}}+
    \frac{TA_{-0}}{1-R^2e^{2i(\omega_1+\Omega)\tau}}\,i
    \biggl[2k_1\Bigl(X_be^{i\Omega\tau}-\sigma_1X_a\Bigr)+
    \delta\Psi_{\textrm{emw}}\biggr],
\end{align*}
\end{subequations}
\end{widetext}
where $\rho_1(\Omega)=R^2e^{2i(\omega_1+\Omega)\tau}$. Phase shift
$\delta\Psi_{\textrm{emw}}$ and factor $\sigma_1$ are introduced in
Sec. \ref{sec_FP_cavity}.


\end{document}